\documentclass[twocolumn,amsmath,amssymb,prl,10pt,nofootinbib,superscriptaddress]{revtex4}
\usepackage{ graphicx, float,amsmath, amsmath, amssymb}
\usepackage[export]{adjustbox}

\def\be{\begin{equation}}
\def\ee{\end{equation}}
\def\bea{\begin{eqnarray}}
\def\eea{\end{eqnarray}}
\def\bse{\begin{subequations}}
\def\ese{\end{subequations}}

\usepackage[breaklinks, colorlinks, citecolor=blue]{hyperref}
\begin{document}
\title{Primordial power spectra from an emergent universe: basic results and clarifications}

\author{Killian Martineau}%
\affiliation{%
Laboratoire de Physique Subatomique et de Cosmologie, Universit\'e Grenoble-Alpes, CNRS/IN2P3\\
53, avenue des Martyrs, 38026 Grenoble cedex, France
}

\author{Aur\'elien Barrau}%
\affiliation{%
Laboratoire de Physique Subatomique et de Cosmologie, Universit\'e Grenoble-Alpes, CNRS/IN2P3\\
53, avenue des Martyrs, 38026 Grenoble cedex, France
}


\date{\today}
\begin{abstract} 
Emergent cosmological models, together with the Big Bang and bouncing scenarios, are among the possible descriptions of the early Universe.
This work aims at clarifying some general features of the primordial tensor power spectrum in this specific framework. In particular, some naive beliefs are corrected. Using a toy model, we investigate the conditions required to produce a scale invariant spectrum and show to which extent this spectrum can exhibit local features sensitive to the details of the scale factor evolution near the transition time. \end{abstract}
\maketitle

\section{Introduction}

The ``Big Bang" term is somehow ambiguous. In a sense, it just refers to the expansion of space, and to the fact that the entire observable universe was, in the past, much smaller, denser and hotter. This is obviously non-controversial. In another sense, it refers to the initial singularity in itself. In this stronger meaning, the very idea of the Big Bang is far from obvious. It is a generic prediction of general relativity (GR) --  remaining usually true in the inflationary paradigm \cite{Borde:1993xh,Borde:1996pt} -- which can, however, be violated in some circumstances.\\

The first important class of models without a Big Bang (in the strong sense) are bouncing models. 
Among the very numerous ways to get a bounce (an excellent review can be found in \cite{Brandenberger:2016vhg}), it is worth mentioning the violation of the null energy condition \cite{Peter:2001fy}, the violation of the strong energy condition \cite{Falciano:2008gt}, the existence of ghost condensates \cite{Lin:2010pf}, galileons \cite{Qiu:2011cy}, S-branes\cite{Kounnas:2011fk}, quintom fields \cite{Cai:2007qw}, higher derivatives \cite{Biswas:2005qr,Biswas:2006bs}, non-standard couplings in the Lagrangian \cite{Langlois:2013cya}, supergravity \cite{Koehn:2013upa}, and loop quantum cosmology \cite{Bojowald:2001xe,Ashtekar:2015dja}. These are only some examples and an exhaustive list should also includes the ekpyrotic and cyclic scenarios \cite{Khoury:2001wf,Steinhardt:2001st},  and, in a way, string gaz cosmology \cite{Battefeld:2005av}. Those ideas are also being investigated in the black hole sector, see \cite{Barcelo:2017lnx} and references therein. \\

The second important class of models beyond the Big Bang are those based on an {\textit{emergent}} scenario. Instead of decreasing and then increasing, the scale factor is, in this case, constant until, at some point, a transition occurs and leads to the current expansion of the Universe. As examples, one can think to (some versions of) nonlinear sigma models \cite{Beesham:2009zw}, Horava-Lifshitz gravity \cite{Wu:2009ah,Mukerji:2010zz}, Einstein-Gauss-Bonnet theory \cite{Mukerji:2010zza}, exotic matter \cite{Paul:2011nw}, branes \cite{Debnath:2011qi}, Kaluza-Klein cosmology \cite{Rudra:2012mu}, particle creation mechanism \cite{Chakraborty:2014ora}, microscopic effects \cite{Perez:2018wlo}, quantum reduced loop gravity \cite{Alesci:2016xqa}, and quintom matter (see \cite{Cai:2012yf} for the background dynamics and \cite{Cai:2013rna} for the associated perturbations). This leads to interesting consequences reviewed for example in \cite{Paul:2014yga,Labrana:2013oca,Zhang:2013ykz,Ghose:2011fk,delCampo:2010kf}.\\

In this article, we focus on emergent models. We don't choose a specific theory but, instead, we try to highlight generic features from a purely phenomenological approach. The aim is not to demonstrate new outstanding results. It simply consists in clarifying the situation, correcting some common misunderstandings and explaining the expected observational features which, to the best of our knowledge, have not been presented so far in a systematic way in the literature. We basically use an ``ad hoc" evolution of the scale factor from a static phase $(a = cte)$ to an inflationary phase $(a\propto e^{H_0t})$ (the subscript 0 does not refer in this context to the value of the Hubble parameter now but to its nearly constant value during inflation). As this transition is expected to be triggered by some event occurring in the evolution of the Universe, we add a small distorsion of the scale factor evolution around the transition time. This distorsion can be a bounce (\textit{i.e} a phase of contraction followed by a phase of expansion) or an anti-bounce (the opposite), that we usually also call a bounce. Both are expected to capture some basic features of emergent models but are also motivated by explicit results obtained, {\it e.g.}, in loop quantum cosmology, or in quantum reduced loop gravity (see the detailed behavior of the scale factor in  \cite{Alesci:2016xqa,Alesci:2018qtm}). The existence of an inflationary stage is natural as soon as a massive scalar field is assumed to be the dominant content of the universe. This will be our implicit hypothesis. In this case, inflation is a strong attractor \cite{Bolliet:2017czc} and occurs nearly inevitably.\\

 We investigate how the primordial tensor power spectrum is affected by variations in the physical characteristics of the features present in the evolution of the scale factor so as to draw a wide picture of the observational characteristics of emergent models. We deliberately decide to focus on tensor perturbations as the scalar spectrum does {\it not} depend only on the scale factor evolution. \\
 
 Throughout all this work, we use Planck units.

\section{Primordial tensor power spectra}

\subsection{The Mukhanov-Sasaki equation for tensor perturbations}

The first order perturbed Einstein equations are equivalent, for a flat FLRW universe and a single matter content modeled by a scalar field, to the gauge-invariant Mukhanov-Sasaki equation:

\begin{equation}
v''(\eta, \vec{x}) -  \vartriangle v(\eta, \vec{x}) - \dfrac{z_{T/S}''(\eta)}{z_{T/S}(\eta)} v(\eta, \vec{x}) = 0 ~.
\label{Mukhanov}
\end{equation}

The ' symbol refers to a derivative with respect to conformal time $\eta$ such that $ad\eta=dt$. This equation depends on two variables $v$ and $z_{T/S}$, called the Mukhanov variables. The canonical variable, $v$, is obtained from a gauge-invariant combination of both the metric coordinate perturbations and the perturbations of the scalar field. The nature of the considered perturbations is encoded in the background variable $z_{T/S}$, in which  the $T/S$ indices refer either to tensor or scalar modes.

Since the background variable writes $z_{S}(t)=a(t)\dot{\Phi}(t)/H(t)$ for scalar modes, $\Phi$ being the scalar field background, the associated evolution highly depends upon the matter evolution. We will therefore not consider scalar perturbations anymore in this study, even if they are currently the most relevant ones for observations. Instead, we will focus on tensor modes, for which the background variable is simply given by $z_{T}(t)=a(t)$. The results and conclusions will therefore be fully generic and usable for any model in  which the scale factor behaves, at least partially, in the way described below, independently of the cause.\\

The Mukhanov-Sasaki equation, that is Eq. (\ref{Mukhanov}), reduces the cosmological evolution of perturbations to the propagation equation of a free scalar field, $v$, with a time-dependent mass $m^{2} = - z_{T}''/z_{T}$ in the Minkowski space-time. The time-dependence of the mass represents the perturbations sensitivity to the dynamical background. 

During the quantization procedure the variable $v$ is promoted to be operator. Its associated Fourier modes satisfy

\begin{equation}
v_{k}''(\eta) + \left(k_{c}^{2} - \dfrac{z_{T}''(\eta)}{z_{T}(\eta)} \right) v_{k}(\eta) =0 ~,
\label{MukhanovSasaki Temporal modes}
\end{equation}

where $k_{c}$ refers to comoving wavenumbers.  This equation can be re-written in cosmic time:

\begin{equation}
    \begin{aligned}
& \ddot{v_{k}}(t) + H(t) \dot{v_{k}}(t)\\
&  + \left( \dfrac{k_{c}^{2}}{a(t)^{2}} - \dfrac{\dot{z}_{T}(t)}{z_{T}(t)}H(t) - \dfrac{\ddot{z}_{T}(t)}{z_{T}(t)} \right) v_{k} =0~~. \\
\Leftrightarrow & \ddot{v_{k}}(t) + H(t) \dot{v_{k}}(t)  + \left( \dfrac{k_{c}^{2}}{a(t)^{2}} - H(t)^{2} - \dfrac{\ddot{a}(t)}{a(t)} \right) v_{k} =0~~.
\end{aligned}
  \label{MukhanovSasakiCosmicTime}
\end{equation}

We introduce a new parameter $h_{k}(t)=v_{k}(t)/a(t)$ such that Eq. (\ref{MukhanovSasakiCosmicTime}) becomes

\begin{equation}
 \ddot{h_{k}}(t)+3H(t)\dot{h_{k}}(t) + \dfrac{k_{c}^{2}}{a(t)^{2}}  h_{k}(t) =0~~.
\label{Mukhanov hk Cosmic Time}
\end{equation}

It is convenient to introduce a second parameter $g_{k}(t) = a(t) \dot{h_{k}}(t)$ in order to rewrite Eq. (\ref{Mukhanov hk Cosmic Time}) as a set of two first order ordinary differential equations (ODEs):

\begin{equation}
\left\{
    \begin{aligned}
& \dot{h_{k}}(t) = \dfrac{1}{a(t)} g_{k}(t)~, \\
& \dot{g_{k}}(t) = - 2 H(t) g_{k}(t) - \dfrac{k_{c}^{2}}{a(t)} h_{k}(t) ~~. 
\end{aligned}
  \right.
  \label{Set EDO Mukhanov}
\end{equation}

\subsection{Initial conditions}

By definition, in the static phase, the scale factor is constant. The propagation equation is then the one of a standard harmonic oscillator,

\begin{equation}
v_{k}''(\eta) + k_{c}^{2} v_{k}(\eta) =0 ~,
\label{initial}
\end{equation}

which can be used to set the usual Bunch-Davies vacuum.  The initial conditions chosen in this work are therefore of the usual type, comparable to what is done in the remote past of a de Sitter state (inflationary model) or in the remote past of a bouncing scenario. Whatever the considered wavenumber, even in the bouncing case, it is always possible to find a time such that the curvature radius can be neglected: the mode effectively ``feels" a Minkowski-like spacetime. As far as initial conditions for the perturbations are concerned, the emergent universe is not different from other usual models. This is true only for tensor modes as the situation is much trickier for scalar ones \cite{Barrau:2018gyz}. 

\section{Purely emergent universe}

\begin{figure}[!h]
\begin{center}
\includegraphics[scale=0.60]{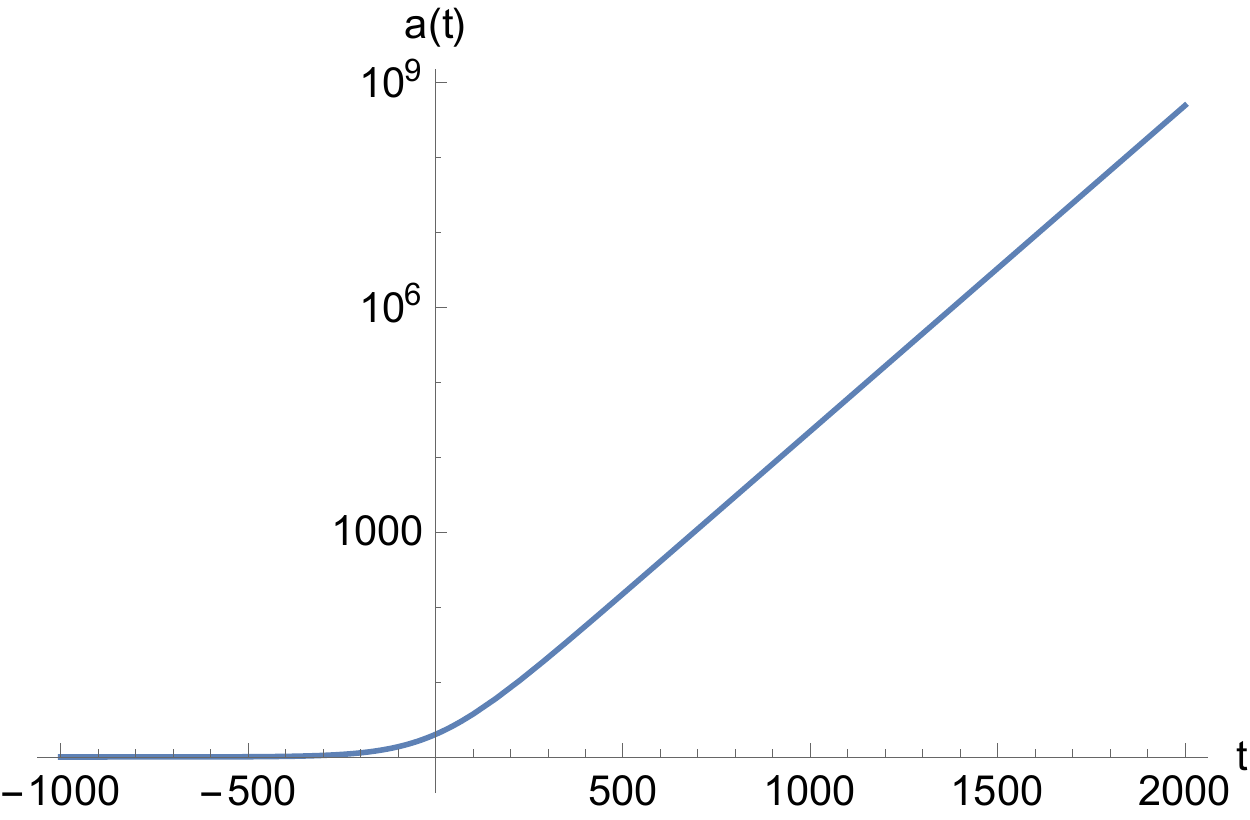}
\caption{Scale factor evolution with $A=1$ and $H_{0}=0.01$.} 
\label{a(t)-no-bounce}
\end{center}
\end{figure}

We model the evolution of an \textit{emergent universe} by a static phase followed by an inflationary stage:

\begin{equation}
a(t) = A + A e^{H_{0} (t-t_{\text{transition}})}~,
\end{equation}

in which $A$ and $H_{0}$ are two constants and $t_{\text{transition}}$ characterizes the time at which the transition between the static and the inflationary phases occurs. If we arbitrarily set $t_{transition}=0$, without any loss of generality, then the scale factor is simply given by $a(t) = A + A e^{H_{0} t}$. The corresponding evolution, with the constants set to $A=1$ and $H_{0}=0.01$, is plotted in arbitrary units in Fig. \ref{a(t)-no-bounce}.\\

Obviously, the constant $A$ in itself has no meaning and can be absorbed in any rescaling of the scale factor. In addition, a modification of the constant in front of the exponential term, such that $a(t) = A + A \mathcal{C}_{1} e^{H_{0} t} ~,~ \mathcal{C}_{1} \in \mathbb{R}^{\star +}$, is simply equivalent to the definition of a new $t_{transition}= -\ln(\mathcal{C}_{1}) / H_{0}$.\\

\begin{figure}[!h]
\begin{center}
\includegraphics[scale=0.40]{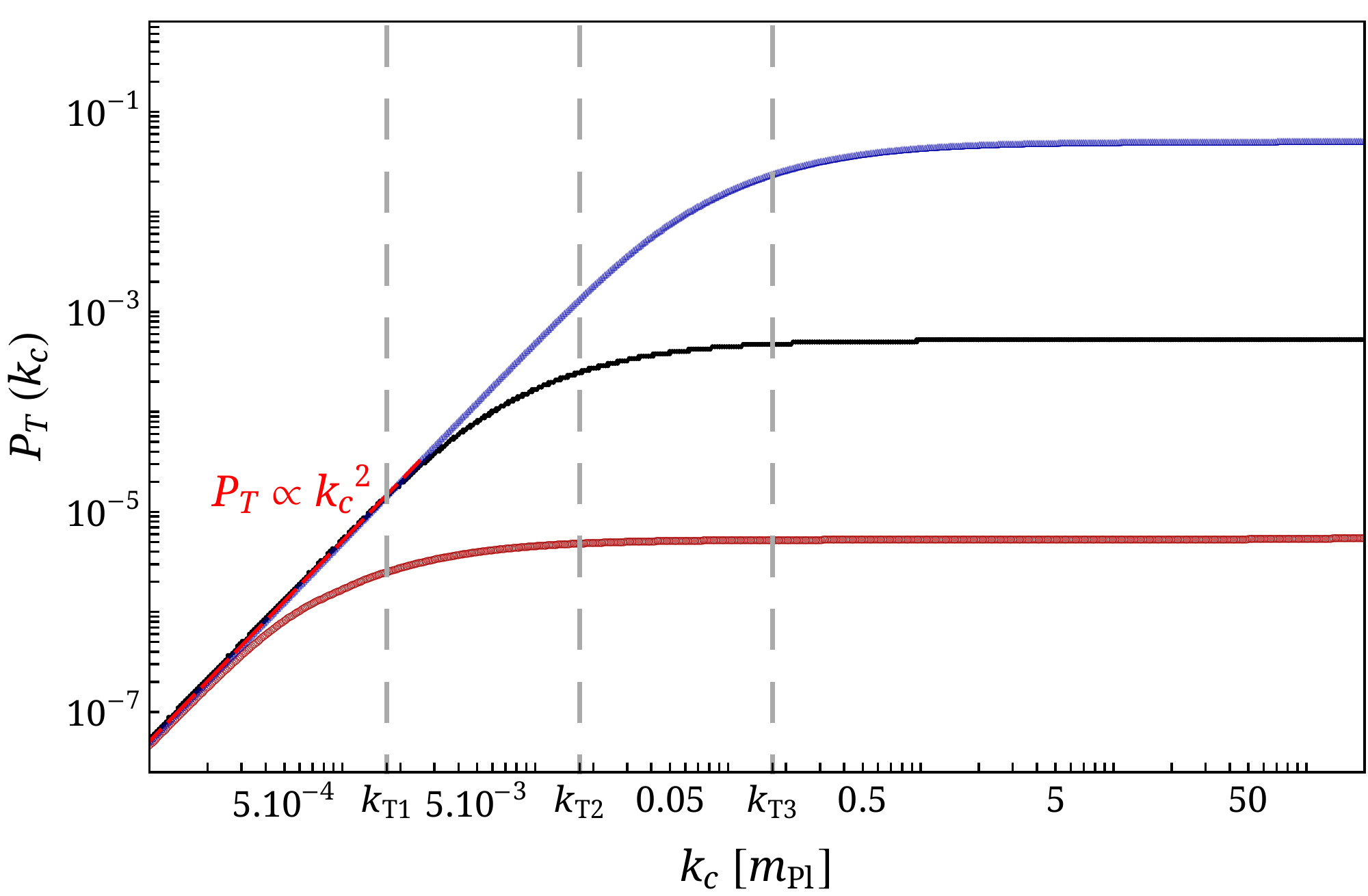}
\caption{Primordial tensor power spectra from the emergent evolution for different values of $H_{0}$. Lower curve (red): $H_{0}=0.001$, mid curve (black): $H_{0}=0.01$ and upper curve (blue): $H_{0}=0.1$.}
\label{spectrum-no-bounce}
\end{center}
\end{figure}

The primordial tensor power spectrum, defined by

\begin{equation}
\mathcal{P}_{T}(k_{c}) = \dfrac{32 k_{c}^{3}}{\pi} \left|\dfrac{v_{k}(t_{e})}{z_{T}(t_{e})} \right|^{2},
\label{tensorspectrum}
\end{equation}

where $t_{e}$ refers to a post-inflationary time (chosen so that the considered modes have exited the horizon), can be explicitly calculated for the evolution of the scale factor given  by Fig. \ref{a(t)-no-bounce}. The result, obtained for different values of $H_{0}$, is shown in Fig. \ref{spectrum-no-bounce}.

Clearly, two regimes do appear in those spectra. First, one can notice a scale-invariant behavior in the ultraviolet (UV), that is for large values of $k_{c}$. Then, a power low behavior appears in the infrared (IR), corresponding to low $k_{c}$ values. The transition scale $k_{T}$ between those two regimes corresponds to the square root of the tensor potential at the transition time, {\it i.e} at $t=0$ in our setting. The tensor potential is given by 

\begin{equation}
\frac{z_{T}''}{z_{T}} = \ddot{a} a + \dot{a}^{2} = A^{2} H_{0}^{2} e^{H_{0} t} + 2 A^{2} H_{0}^{2} e^{2 H_{0} t}
\end{equation}

and its value at the transition is then

\begin{equation}
\left. \frac{z_{T}''}{z_{T}}\right|_{t=0} = 3 A^{2} H_{0}^{2}~~.
\end{equation}

For example, in the case $H_{0}=0.01$ displayed in the mid curve of Fig. \ref{spectrum-no-bounce}, the transition scale is $k_{T2} = \sqrt{\left. \frac{z_{T}''}{z_{T}}\right|_{t=0}} = \sqrt{3.10^{-4}} \simeq 1.7\times10^{-2}$. The  dependence of the transition scale upon $H_{0}$ also appears clearly since $k_{T1} \simeq 1.7\times10^{-3}$ and $k_{T3} \simeq 1.7\times10^{-1}$.\\

 This already raises two basic points. First, the naive view according to which the causal contact made possible by the static phase, where $H=0$ ($R_H\to \infty$), would be sufficient to ensure a spectrum compatible with observation is obviously wrong. Inflation (or other processes leading to scale invariance) is still needed. Second, the way inflation begins does matter and sets the scale above which the spectrum becomes (nearly) flat.

\section{Emergent universe with a bounce}

The previously considered situation is clearly over-simplified. We now make the model slightly more complicated by adding a ``feature" in the evolution of the scale factor before the transition to the inflationary period. As  mentioned in the previous section, the interesting -- and somehow usually under-estimated -- fact about emergent models is that the spectrum does depend on the details of the transition period. Some informations on this specific period might therefore be observationally attainable. In addition, some concrete models of quantum gravity lead to a ``mini-bounce" before the transition. This is, for example, the case in quantum reduced loop gravity \cite{Alesci:2016xqa,Alesci:2018qtm}. This model was designed to consistently study symmetry reduced systems within the loop quantum gravity framework (see, {\it e.g.} \cite{Rovelli:2011eq}). In particular, it bridges the gap between effective cosmological models of loop quantum cosmology \cite{lqc9} and the full theory, addressing the dynamics before any minisuperspace reduction \cite{Alesci:2014rra}. This basically preserves the graph structure and SU(2) quantum numbers. It was explicitly shown that this model leads to a little bounce (or even to several mini-bounces) preceding the inflationary stage. Beyond this specific case, one can generically expect a footprint in the evolution of the scale factor of whatever physical phenomenon has triggered the transition. In the following, we therefore perturb the scale factor evolution just before the inflationary stage to study how the primordial tensor power spectrum is sensitive to the details of this distortion.\\

The scale factor evolution is now modeled by the following function:

\begin{eqnarray}
& & a(t)  =  A + A e^{H_{0} t} + \frac{A \times C}{\arctan\left(B_{1} \sigma_{1}\right)-\arctan\left(B_{2} \sigma_{2}\right)} \times \\ \nonumber
&& \left\lbrace \arctan \left[ B_{1} \left( t-\left( \mu - \sigma_{1} \right) \right) \right] - \arctan\left[B_{2} \left(t-\left(\mu - \sigma_{2} \right) \right) \right] \right\rbrace .
\end{eqnarray}

The constant $C$ characterizes the bounce amplitude, $\mu$ is its mean value, $\sigma_{1}$ and $\sigma_{2}$ allow to set the width, and $B_{1}$ and $B_{2}$ correspond to the steepness. 
The term $\left[\arctan\left( B \sigma_{1}\right)-\arctan\left(B \sigma_{2}\right)\right]^{-1}$ is just a normalization to ensure that the bounce amplitude remains constant under variations of $B$, $\sigma_{1}$ and $\sigma_{2}$. In the following, we set $B_{1} = B_{2} = B$, to focus on symmetrical bounces. The influence of an asymmetry is a higher order effect which is beyond the scope of this study. We also choose $\sigma_{1} = - \sigma_{2} = \sigma$.
The scale factor is finally expressed as

\begin{eqnarray}
& & a(t)  =  A + A e^{H_{0} t} + \frac{A \times C}{2 \arctan\left(B \sigma \right)} \times \\ \nonumber
&& \left\lbrace \arctan \left[ B \left( t-\left( \mu - \sigma \right) \right) \right] - \arctan\left[B \left(t-\left(\mu + \sigma \right) \right) \right] \right\rbrace .
\end{eqnarray}

\begin{figure}[!h]
\begin{center}
\includegraphics[scale=0.53]{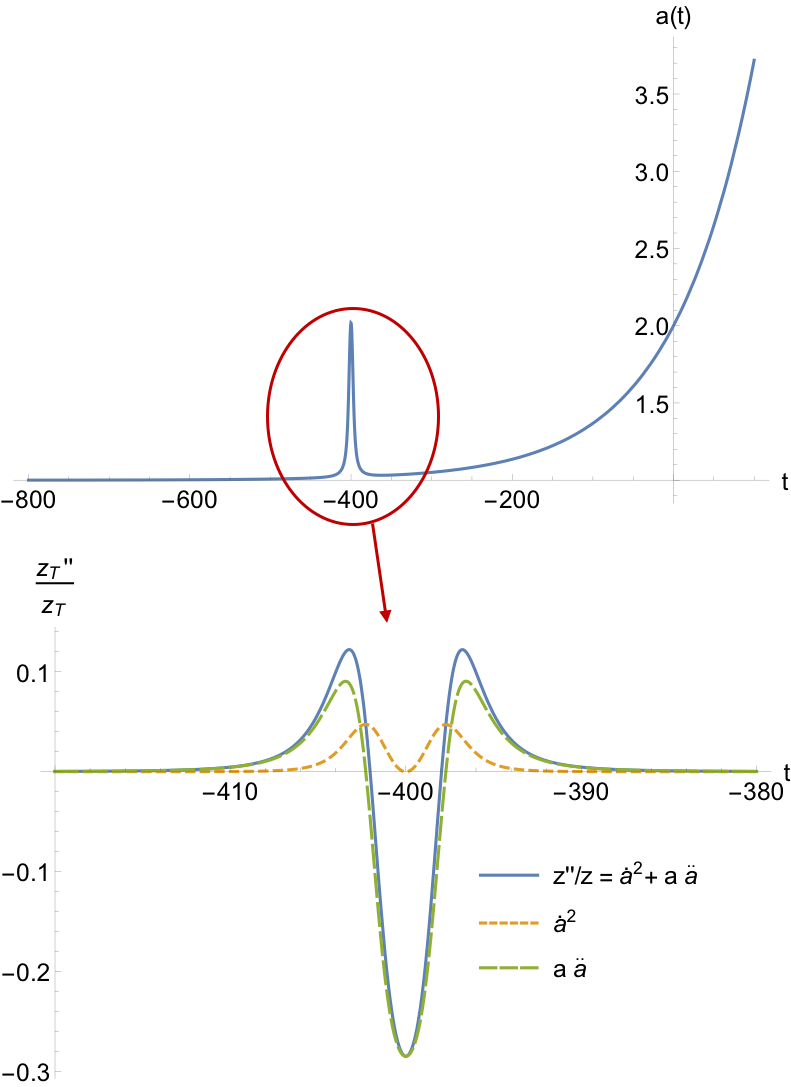}
\caption{\underline{Upper panel:} Scale factor evolution with one bounce characterized by  $\mathcal{C}_{1} = 1$, $\mu =-400$, $\sigma = 2$ and $B=0.4$. \underline{Lower panel:} Tensor potential around the bounce.} 
\label{one bounce scale factor and potential}
\end{center}
\end{figure}

Arbitrarily choosing $A=1$ and $H_{0}=10^{-2}$, as in the case without any bounce, and fixing $\mathcal{C} = 1$, $\mu =-400$, $\sigma = 2$ and $B=0.4$, the scale factor evolution is displayed in the upper panel of Fig. \ref{one bounce scale factor and potential}. The lower panel  shows the associated tensor potential around the bounce. It is worth noticing that the ``sign" of the bounce has no influence on the spectrum. It is displayed on Fig. \ref{one bounce scale factor and potential} as a local increase of the scale factor, but should we choose the other sign, leading to a decrease of the scale factor, the spectrum would remain the same as it will be shown later.

\begin{figure}[!h]
\begin{center}
\includegraphics[scale=0.40]{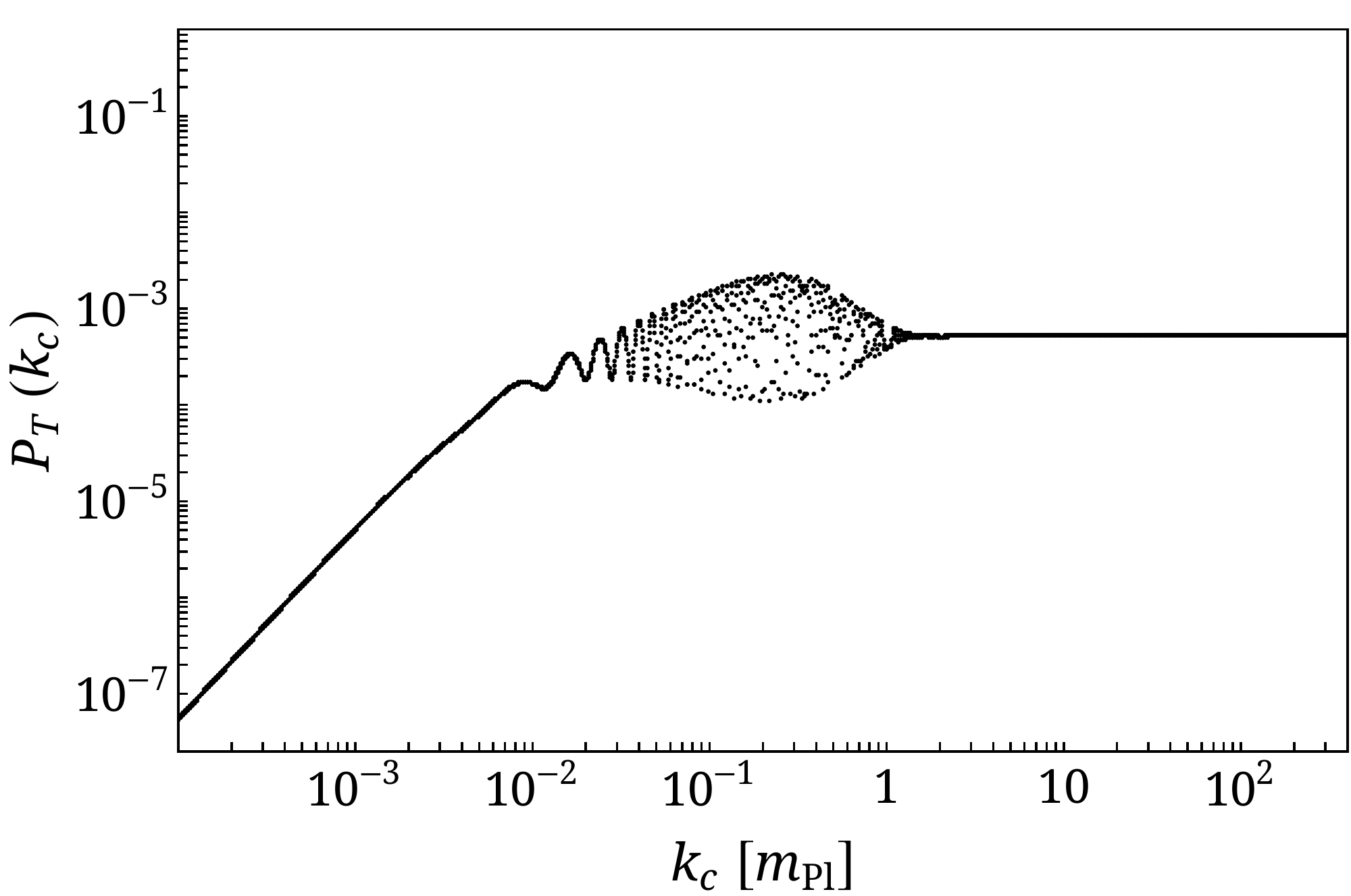}
\caption{Primordial tensor power spectrum obtained from the scale factor evolution with one bounce characterized by  $\mathcal{C} = 1$, $\mu =-400$, $\sigma = 2$ and $B=0.4$.} 
\label{spectrum of reference} 
\end{center}
\end{figure}

The primordial tensor power spectrum computed with this background evolution is given Fig. \ref{spectrum of reference}. 

First, one can notice that the general trend is the same as in the case without bounce, which is not surprising as the main tendencies are driven by the choice of the initial state and the existence of an inflationary stage. The spectrum is still scale-invariant in the UV and grows proportionally to $k_{c}^{2}$ in the IR. However, the bounce does have an impact on the spectrum: it induces oscillations in the $k_{c}$ space. The envelope of the oscillations forms a kind of ``bullet" in the spectrum. Those oscillations can be traced back to the time evolution of the mode functions which becomes highly $k_c$-dependent in the presence of a bounce. This clearly establishes an observational window on the detailed behavior of the Universe close to the emergent time (and even before). This also contradicts a second naive belief according to which whatever happens before inflation is washed out by inflation. The details of the transition regime might be observationally probed.\\

This spectrum will be the reference one for the rest of this study. We now investigate how it depends (amplitude, IR and UV limits, shape of the ``bullet", etc.) on the different parameters of the model.\\

\begin{figure}[!h]
\begin{center}
\includegraphics[scale=0.65]{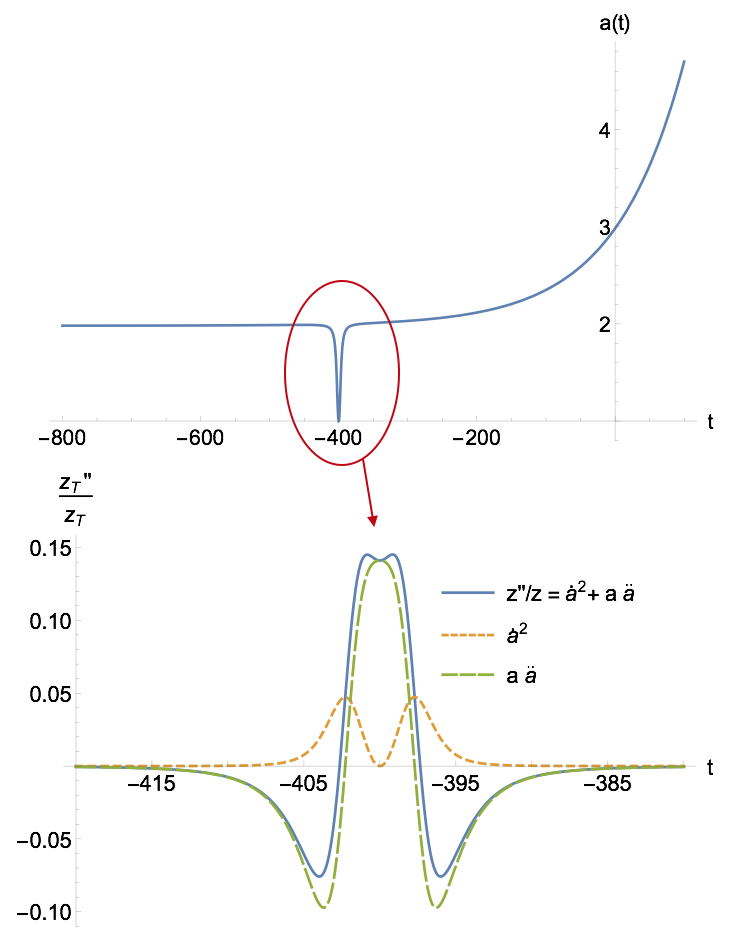}
\caption{\underline{Upper panel:} Scale factor evolution with one bounce of negative sign characterized by $\mathcal{C} = 1$, $\mu =-400$, $\sigma = 2$ and $B=0.4$. \underline{Lower panel:} Tensor potential around the bounce. } 
\label{one bounce scale factor down}
\end{center}
\end{figure}

\begin{figure}[!h]
\begin{center}
\includegraphics[scale=0.40]{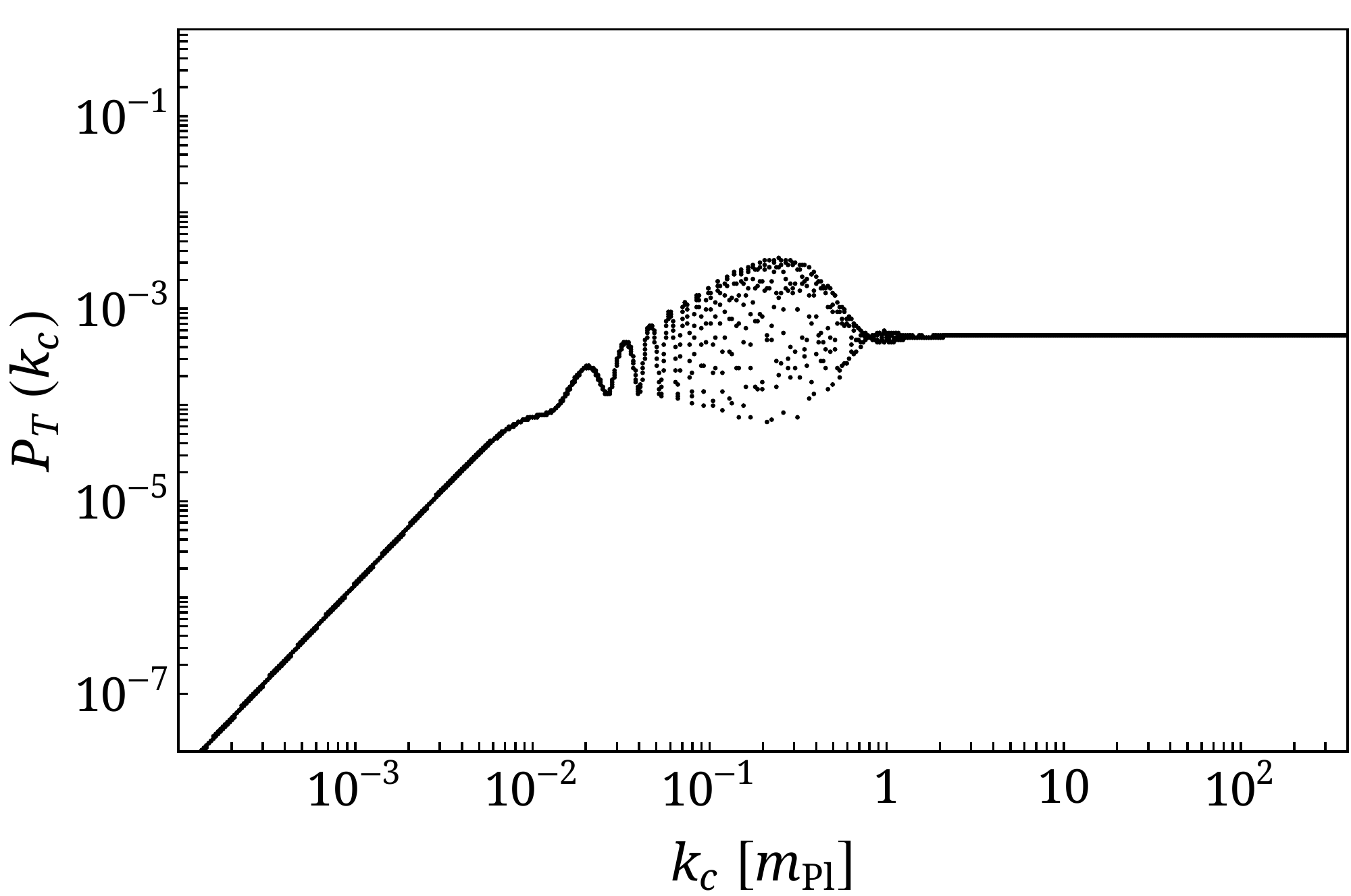}
\caption{Primordial tensor power spectrum associated with the scale factor evolution with one bounce of negative sign characterized by $\mathcal{C} = 1$, $\mu =-400$, $\sigma = 2$ and $B=0.4$.} 
\label{spectrum neg}
\end{center}
\end{figure}

As previously mentioned, all the results derived in this work remain valid if the bounce is of negative sign, that is corresponds to a transition between a (locally) contracting and an expanding phase. A negative bounce of this kind is shown in Fig. \ref{one bounce scale factor down}, together with the corresponding potential. The resulting spectrum in displayed on Fig. \ref{spectrum neg} and can hardly be distinguished from the reference one. 

\subsection{Impact of the bounce parameters on the primordial tensor power spectra}

The aim of this section is to study how variations of the bounce parameters $B$, $C$, $\mu$ and $\sigma$ modify the shape of the primordial tensor power spectrum. Even if those oscillations cannot be currently observed, it is still interesting to see if general trends appear. Many experiments are being operated or considered to measure B-modes in the cosmological microwave background (CMB).  Since the  toy-model presented in this article is basically independent of the details of the quantum cosmology or modified gravity theory considered (as long as the Mukhanov-Sasaki equation remains valid), the results presented can easily be applied or adapted to forthcoming emergent, bouncing or emergent-bouncing cosmological models.

\subsubsection{\textbf{The position of the bounce $\mu$}}

\begin{figure}[!h]
\begin{center}
\includegraphics[scale=0.40]{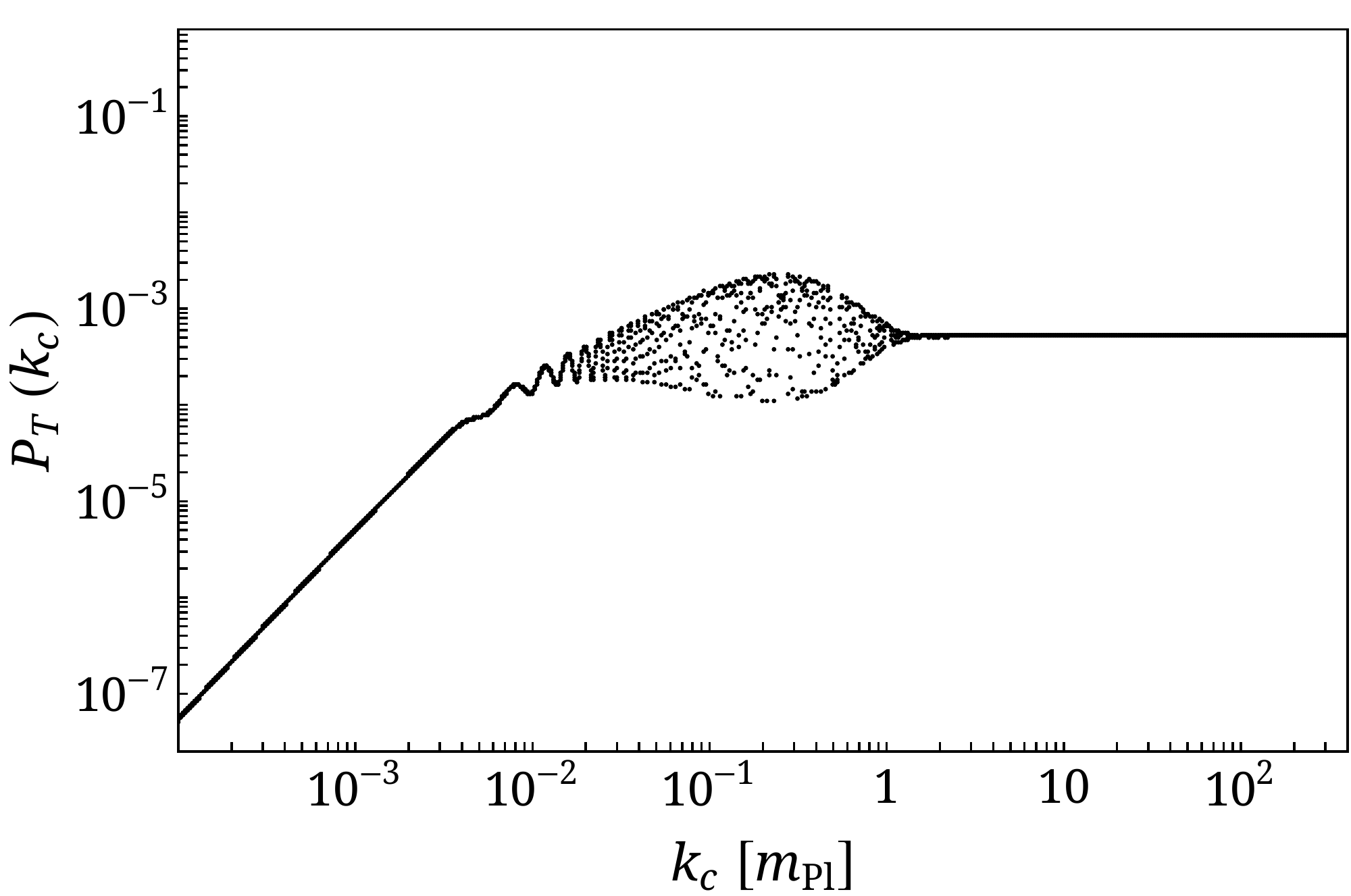}
\caption{Primordial tensor power spectrum associated to an evolution with a bounce identical to the reference case, \textit{i.e} $\mathcal{C} = 1$, $\sigma = 2$ and $B=0.4$, but shifted to $\mu=-800$.} 
\label{bounce shifted}
\end{center}
\end{figure}

First, let us study whether the position of the bounce in the static phase plays an important role in the characteristics of the spectrum. To this aim, it is enough to change the value of $\mu$. It appears that the bounce position in the static phase has almost no consequence on the primordial tensor power spectrum. For example, Fig. \ref{bounce shifted} shows the spectrum obtained for a bounce similar to the reference one, but shifted to $\mu=-800$. It can easily be seen that this spectrum is very close to the reference one. The numerical investigations show, beyond this particular example, that the position of the bounce has no significant influence on the spectrum whatever its position in the static phase. This, in principle, opens an observational window on arbitrarily remote times in the history of the Universe. Once the ``instability" is triggered, the time at which it takes place if basically of no relevance. 

\subsubsection{\textbf{The steepness of the bounce $B$}}

\begin{figure}[!tbp]
  \begin{minipage}[t]{0.4\textwidth}
  \hspace{1.3 cm}
\includegraphics[scale=0.45]{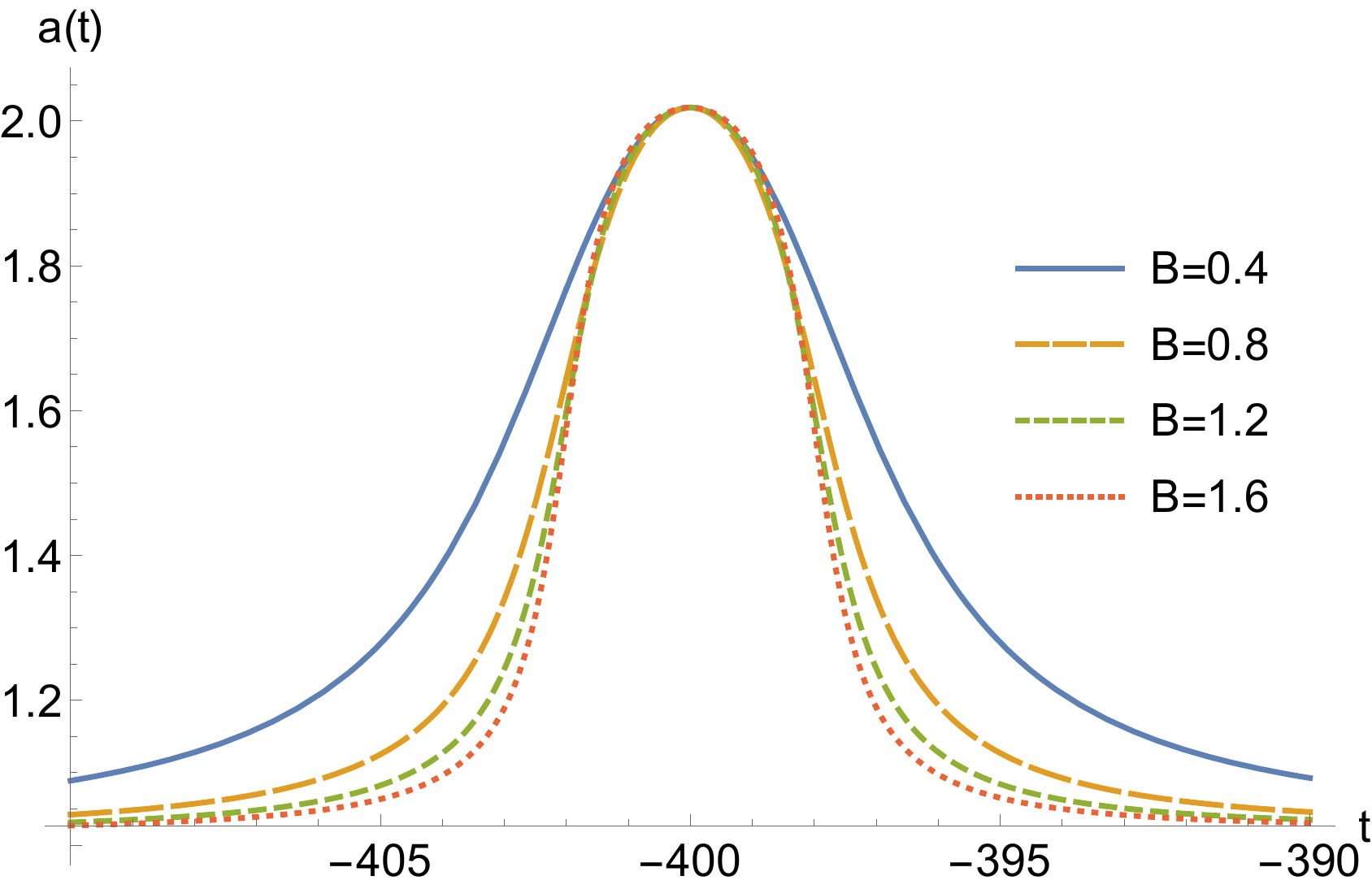}
\label{b values}
  \end{minipage}
  \vspace{0.2 cm}
  \begin{minipage}[t]{0.4\textwidth}
\includegraphics[scale=0.45]{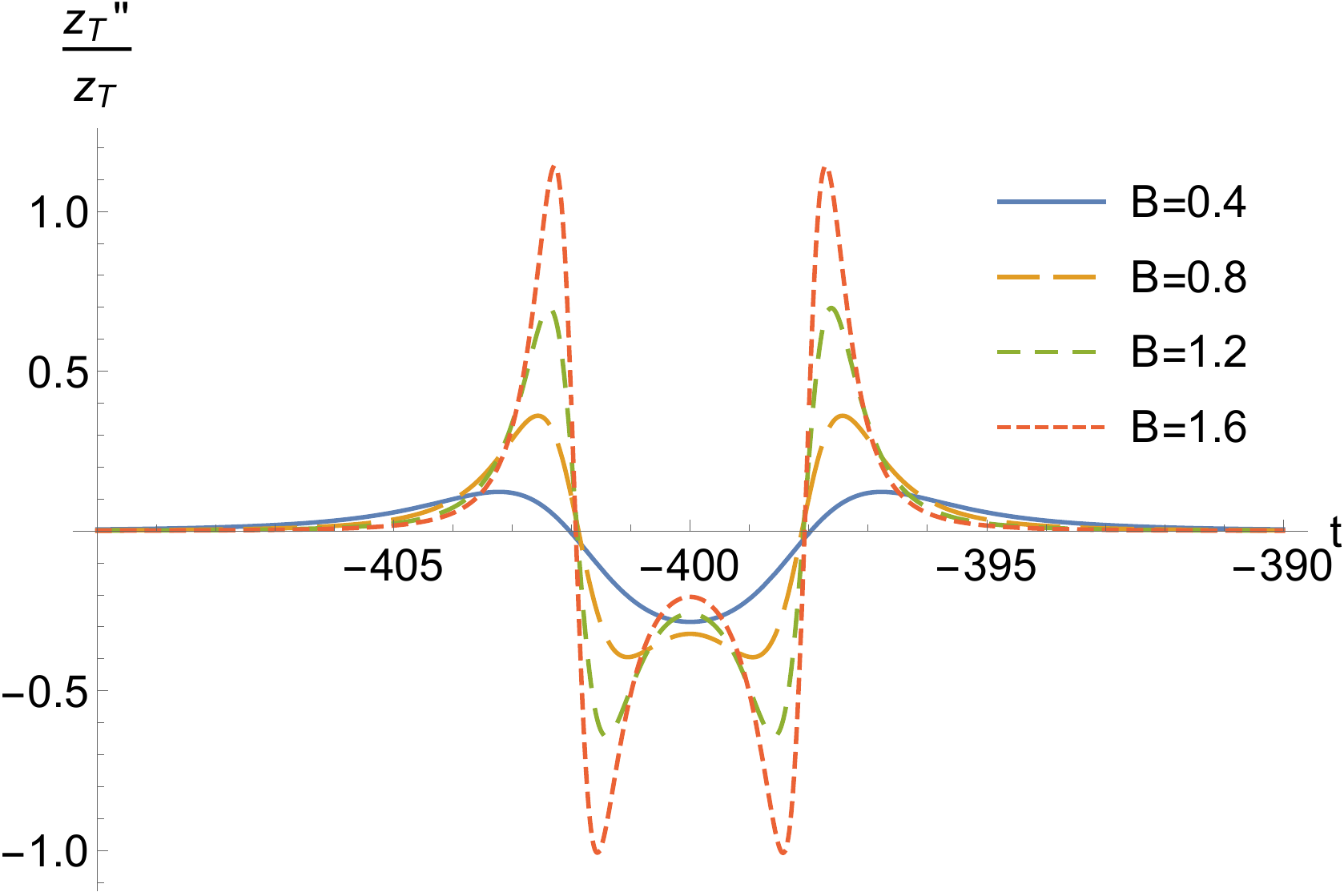}
\label{potential b values}
  \end{minipage}
   \caption{\underline{Upper panel:} Evolution of the scale factor with $\mathcal{C} = 1$, $\mu =-400$, $\sigma = 2$ but different values of $B$. \underline{Lower panel:} Associated tensor potentials around the bounce.}  
   \label{b values plot}
\end{figure}

To study the impact of the bounce steepness, \textit{i.e} its ``slope", we vary the parameter $B$.
The bounce dependence upon this parameter is presented in  Fig. \ref{b values plot}, together with the associated tensor potentials. Since the tensor potential is highly sensitive to variations of the bounce steepness (as it includes derivatives of the scale factor), only small variations of $B$ are represented.\\

\begin{figure}[!tbp]
  \begin{minipage}[t]{0.4\textwidth}
  \hspace{1.3 cm}
\includegraphics[scale=0.35]{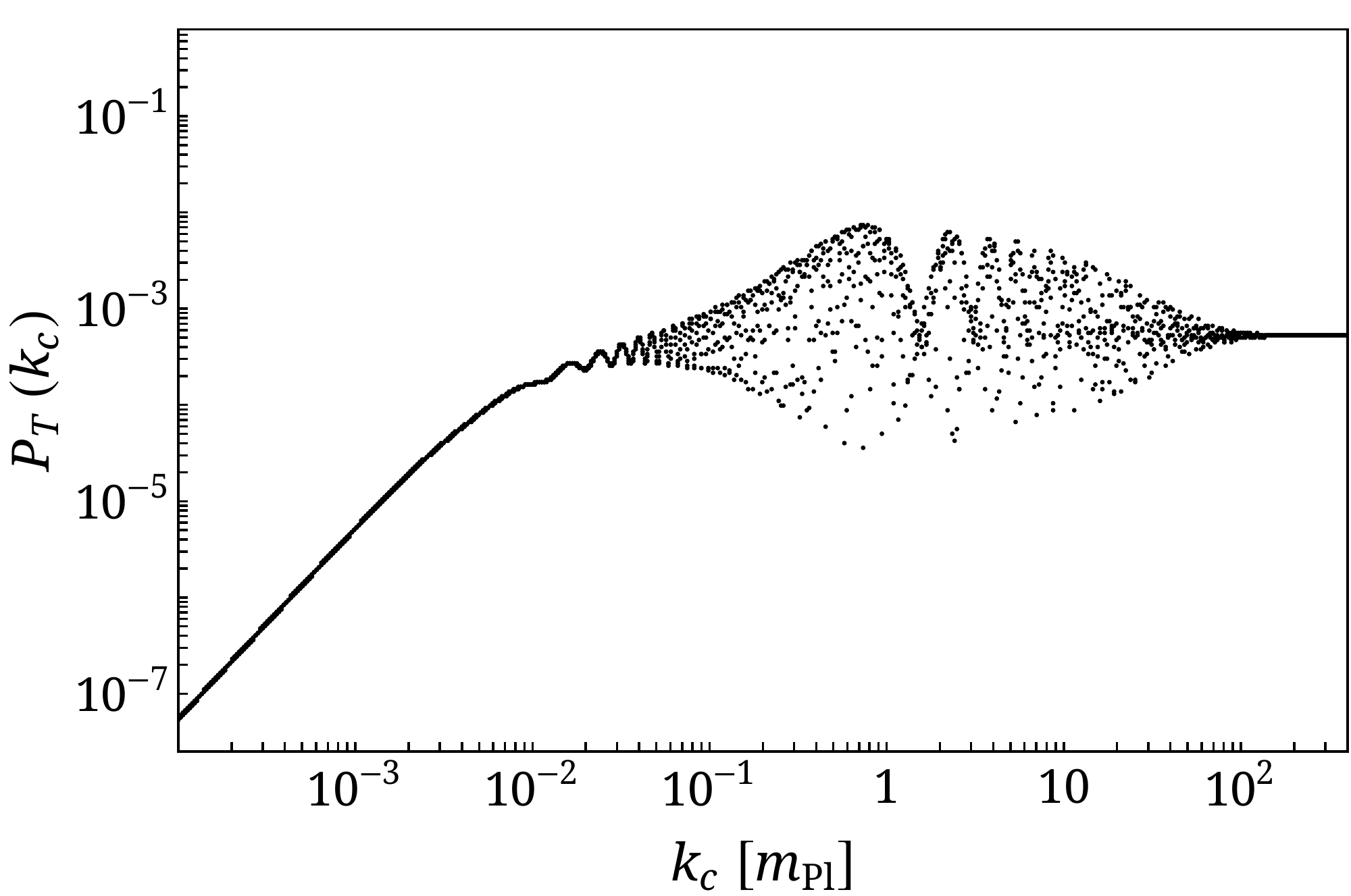}
\label{spectrum B=40 sigma=2}
  \end{minipage}
  \vspace{0.2 cm}
  \begin{minipage}[t]{0.4\textwidth}
\includegraphics[scale=0.35]{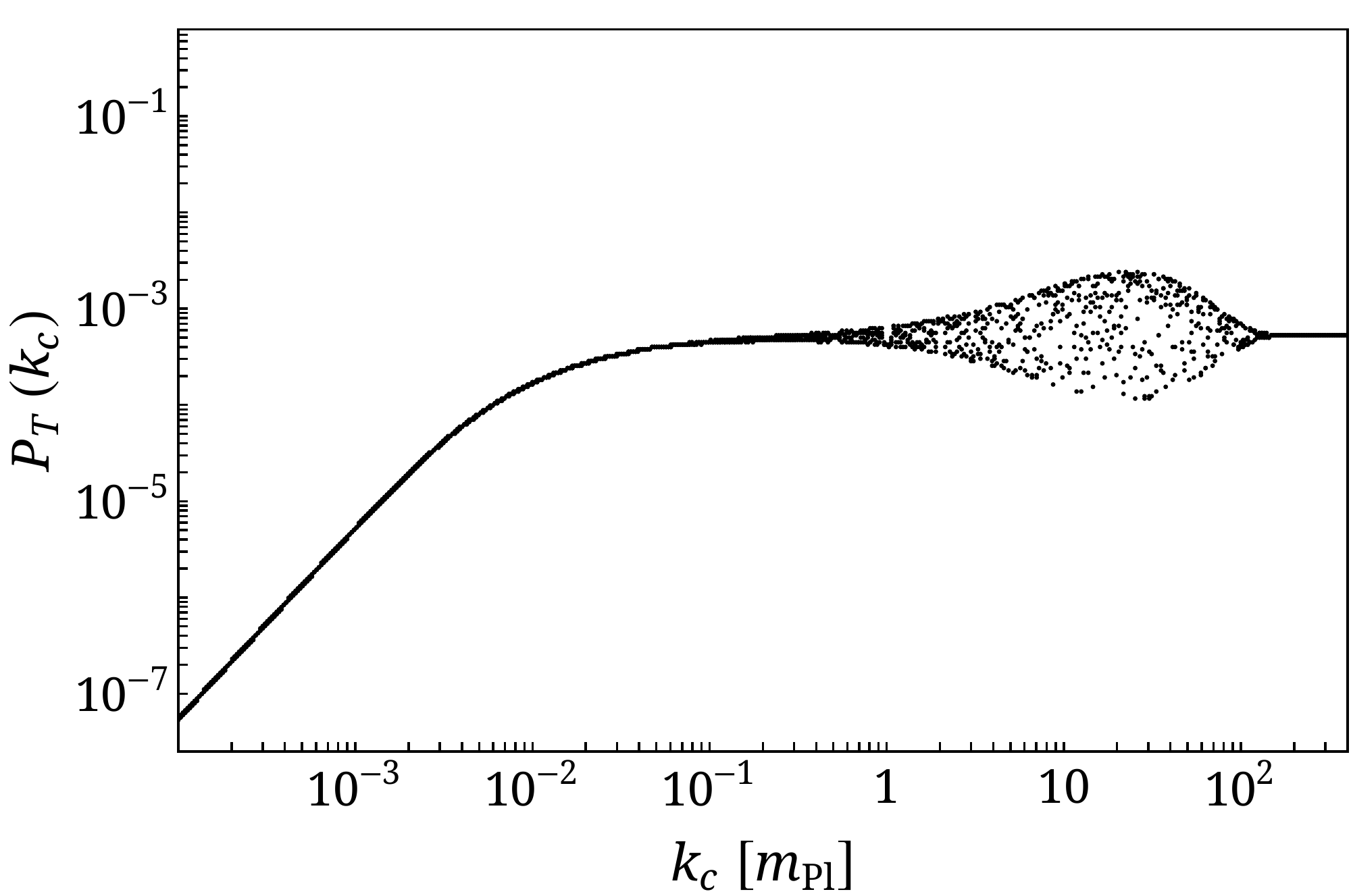}
\label{spectrum B=40 sigma=0,02}
\end{minipage}
\caption{\underline{Upper panel:} Primordial tensor power spectrum obtained with a steep bounce characterized by $\mathcal{C} = 1$, $\mu =-400$, $\sigma = 2$ and $B=40$. \underline{Lower panel:} Spectrum with a narrower bounce (than in the upper panel): $\sigma = 0.02$.} 
   \label{spectra B=40}
\end{figure}

The larger the value of $B$, the steeper the bounce. The (local) maximum value of the potential at the bounce thus increases with $B$. We therefore expect that the range of $k_{c}$  corresponding to modes impacted by (or sensitive to) the bounce is shifted to higher $k_{c}$ values compared to the reference case.
Let us consider a cosmic evolution where the bounce has been highly steepened when compared to the reference case. We choose $B=40$ (one hundred times higher than the value of the reference case), the values of the other parameters being the same as in the reference case. The resulting power spectrum is shown in the upper panel of Fig. \ref{spectra B=40}. The two frequencies appearing in the plot are associated width the two scales of the problem (width of the bounce and rise time of the edge).\\

The size of the ``bullet" (or range of oscillations) in the $k_{c}$ space is extended up to higher values. This is an interesting point as the ``low"-$k_c$ features are often considered to be hard to be experimentally probed. For exemple, in loop quantum cosmology, deviations from scale invariance, in the form of oscillations, happen in the IR (see \cite{Ashtekar:2009vc,Bolliet:2015bka}). They are often considered as extremely difficult to probe as this would require a very high level of fine-tuning. The comobile values of the wavenumbers that can be seen in the CMB are actually set by the duration (number of e-folds) of inflation. In loop quantum cosmology, the interesting IR features can only be seen if this number is arbitrarily set to its lowest experimentally allowed value \cite{Barrau:2016nwy}. This makes the model difficult to be probed unless new specific features appear in the UV, {\it e.g.} through trans-planckian effects \cite{Martineau:2017tdx} or because of a change of signature \cite{Schander:2015eja}. The effect underlined here, that is the displacement or widening of the ``bullet" to larger values of $k_c$ because of the steepness of the mini-bounce, is therefore of potential observational significance. The specific features might be probed without fine-tuning the number of inflationary e-folds to its lowest allowed value (around $N\sim 60$). It is worth reminding that, in principle, if the evolution starts at the Planck density and if the Universe is filled with a massive scalar field, the number of e-fold can be anything between 0 and a few $10^{14}$ (and remains compatible with observations ). In some bouncing cases this number of e-folds can be predicted \cite{bl,Martineau:2017sti} by the model but this remains an open issue for emergent scenarios (finding a known probability distribution function for initial conditions is trick unless the existence of an oscillating phase for the field is demonstrated).  

The lower panel of Fig. \ref{spectra B=40} corresponds to a reduction of the width of the bounce (by factor one hundred) with respect to the previous case. The shape of the distortion gets closer to the reference one but, as expected, the ``bullet" is translated toward the higher $k_{c}$ regime.\\

Obviously, it the bounce is smoothed (by a decrease of $B$), the maximum of the tensor potential decreases and the opposite effect occurs: the oscillations are shifted to the IR regime, which  is far less interesting for phenomenology. \\




A modification of the steepness of the bounce -- presumably associated with the triggering of the transition from the static to the inflationary phase -- has a strong impact on the shape of the tensor potential at the bounce. The range of comoving modes sensitive to the bounce thus highly depends on the steepness of the evolution of the scale factor. This establishes that, as far as phenomenology is concerned, a very steep bounce is more likely to be observable, even if it occurs in most remote past of the Universe. 

\subsubsection{\textbf{The amplitude of the bounce $C$}}

\begin{figure}[!tbp]
  \begin{minipage}[t]{0.4\textwidth}
\includegraphics[scale=0.50]{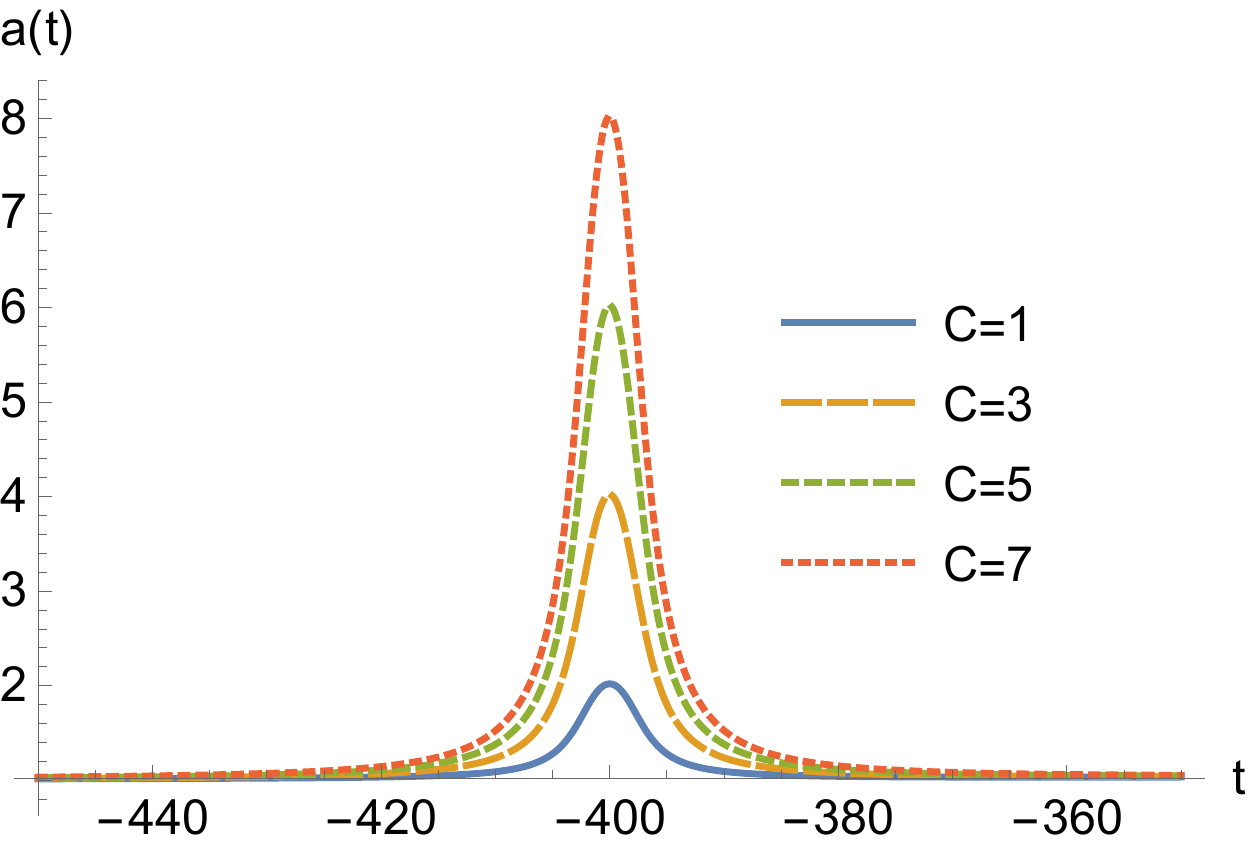}
\label{C values}
  \end{minipage}
  \vspace{0.2 cm}
  \begin{minipage}[t]{0.4\textwidth}
\includegraphics[scale=0.50]{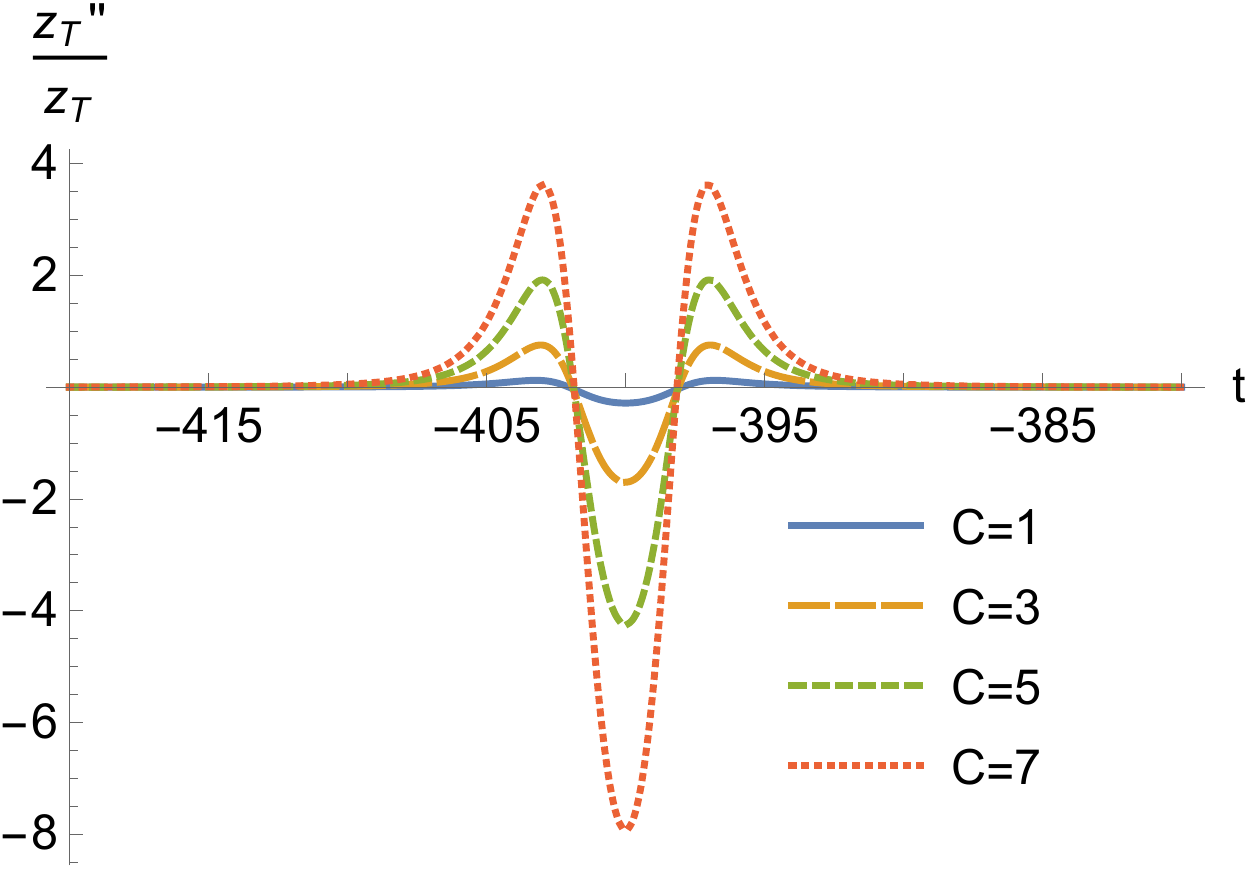}
\label{potential C values}
  \end{minipage}
   \caption{\underline{Upper panel:} Scale factor evolution around the bounce. From bottom to top : increasing values of $\mathcal{C}$. \underline{Lower panel:} Associated tensor potentials.} 
   \label{C values plot}
\end{figure}

We now study the consequences of variations of the bounce amplitude on the primordial tensor spectrum. Figure \ref{C values plot} displays, in the upper panel, the effect of a variation of the factor $C$ entering the scale factor evolution. The lower panel shows the associated potentials.

The amplitude of the bounce in itself has no meaning. The important parameter is the ratio between the extremal value of the scale factor at the bounce and its value in the static phase. This is the relevant parameter which is varied. \\

\begin{figure}[!tbp]
  \centering
  \begin{minipage}[t]{0.4\textwidth}
   \includegraphics[scale=0.37]{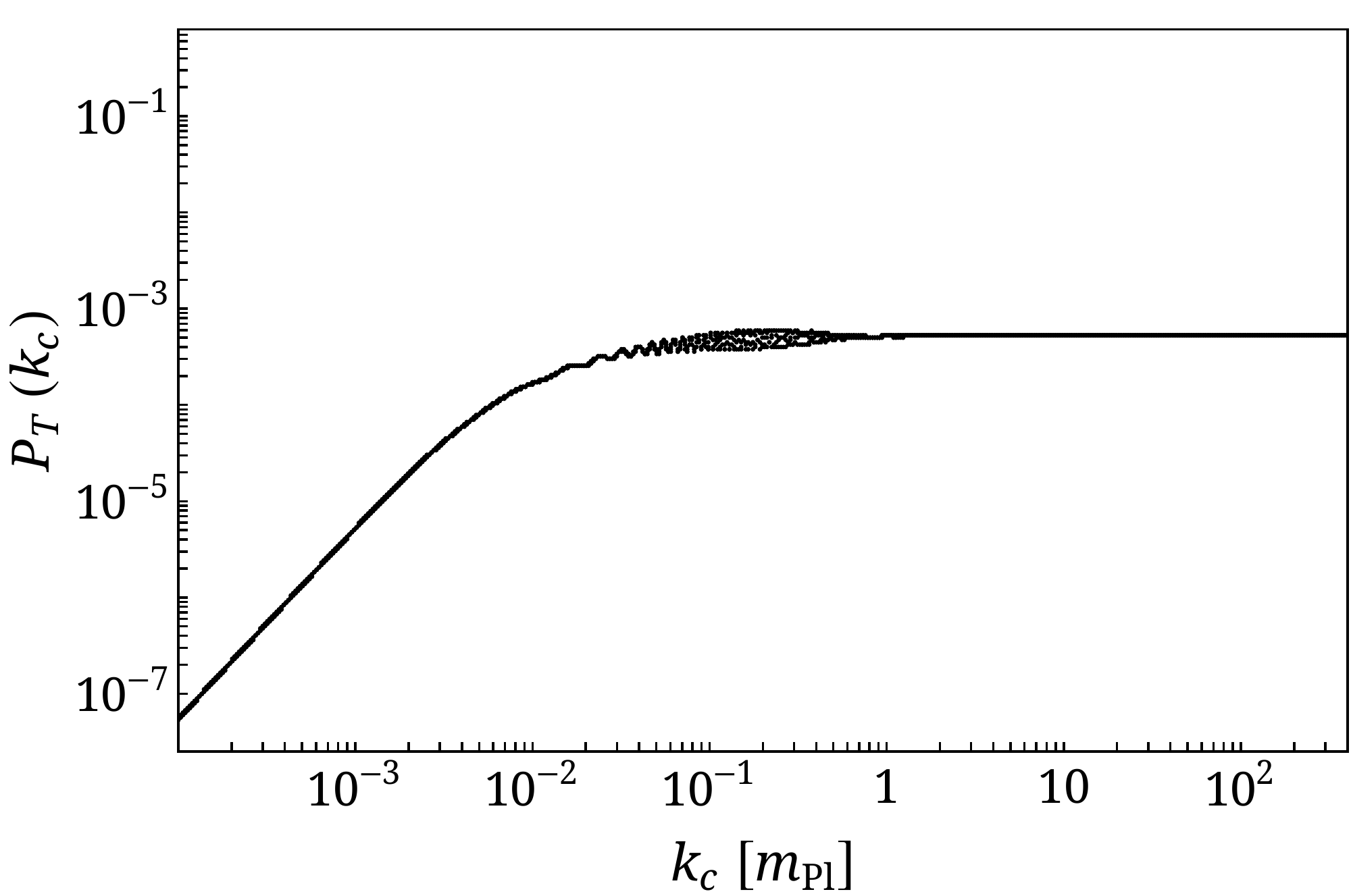}
   \label{C=0.1}
  \end{minipage}
  \begin{minipage}[t]{0.4\textwidth}
     \includegraphics[scale=0.37]{referenceSpectrum.pdf}
     \label{C=1}
  \end{minipage}
  \begin{minipage}[t]{0.4\textwidth}
     \includegraphics[scale=0.37]{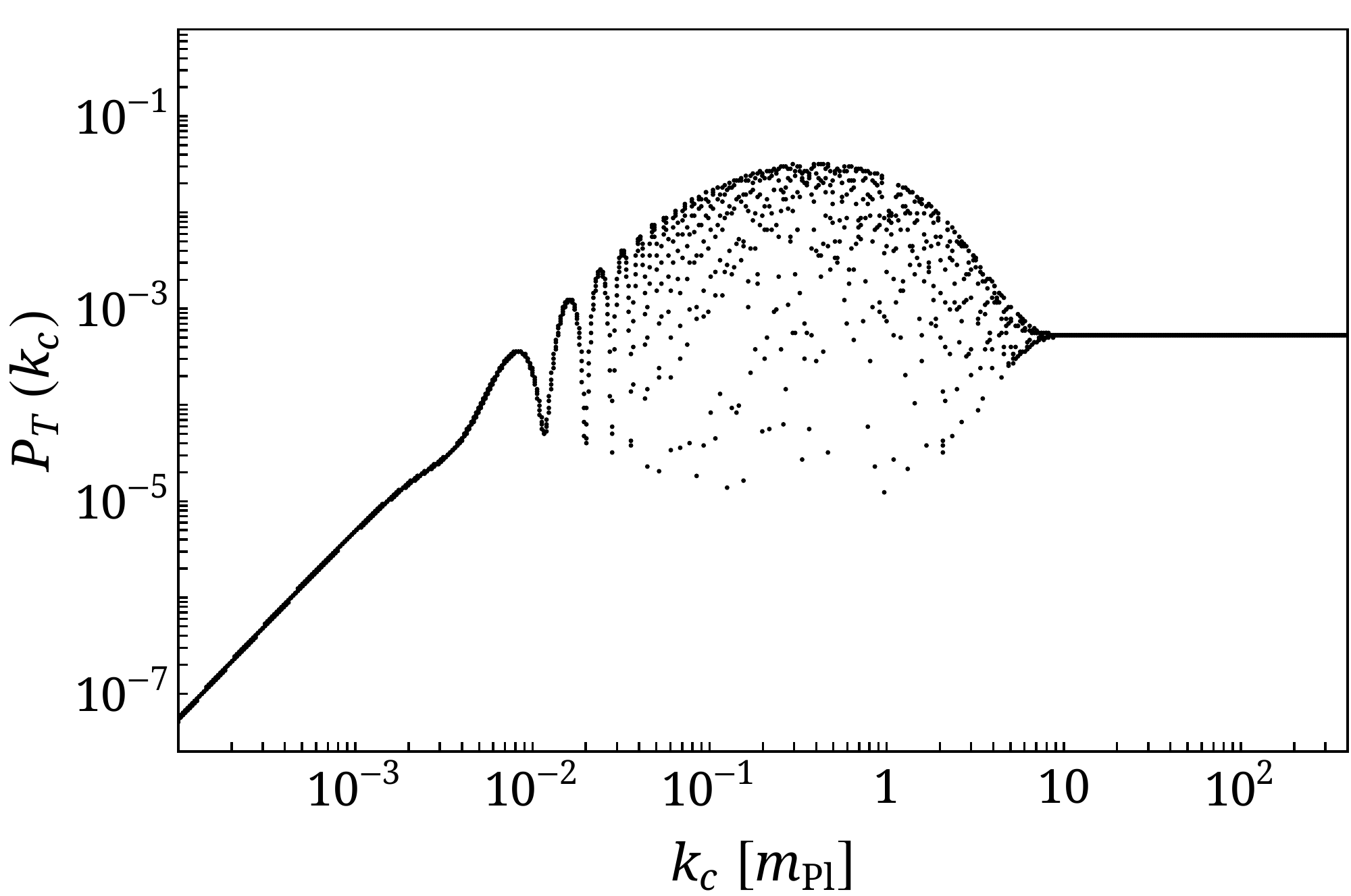}
     \label{C=10}
  \end{minipage}
   \caption{Primordial tensor spectra for different values of the bounce amplitude $\mathcal{C}$, the other parameters being unchanged with respect to the reference case. \underline{Upper panel:} $\mathcal{C}=0.1$, \underline{mid panel:} $\mathcal{C}=1$ (reference case), \underline{Lower panel:} $\mathcal{C}=10$.}  
   \label{Spectra different C}
\end{figure}

The primordial tensor power spectra, for $A$, $H_{0}$, $B$ and $\sigma$ taken as in the reference case but for different values of the bounce amplitude, given by $C=0.1$, $C=1$ (reference case), and $C=10$, are shown in Fig. \ref{Spectra different C}. It can easily be seen that an increase in the amplitude of the bounce amplifies the oscillations in the $k_{c}$ space. This both opens a possible observational window and allows, in principle, to put constraints on the amplitude of the bounce using upper limits on the tensor-scalar ratio.

\subsubsection{\textbf{The width of the bounce $\sigma$}}

\begin{figure}[!tbp]
  \centering
  \begin{minipage}[t]{0.4\textwidth}
   \includegraphics[scale=0.40]{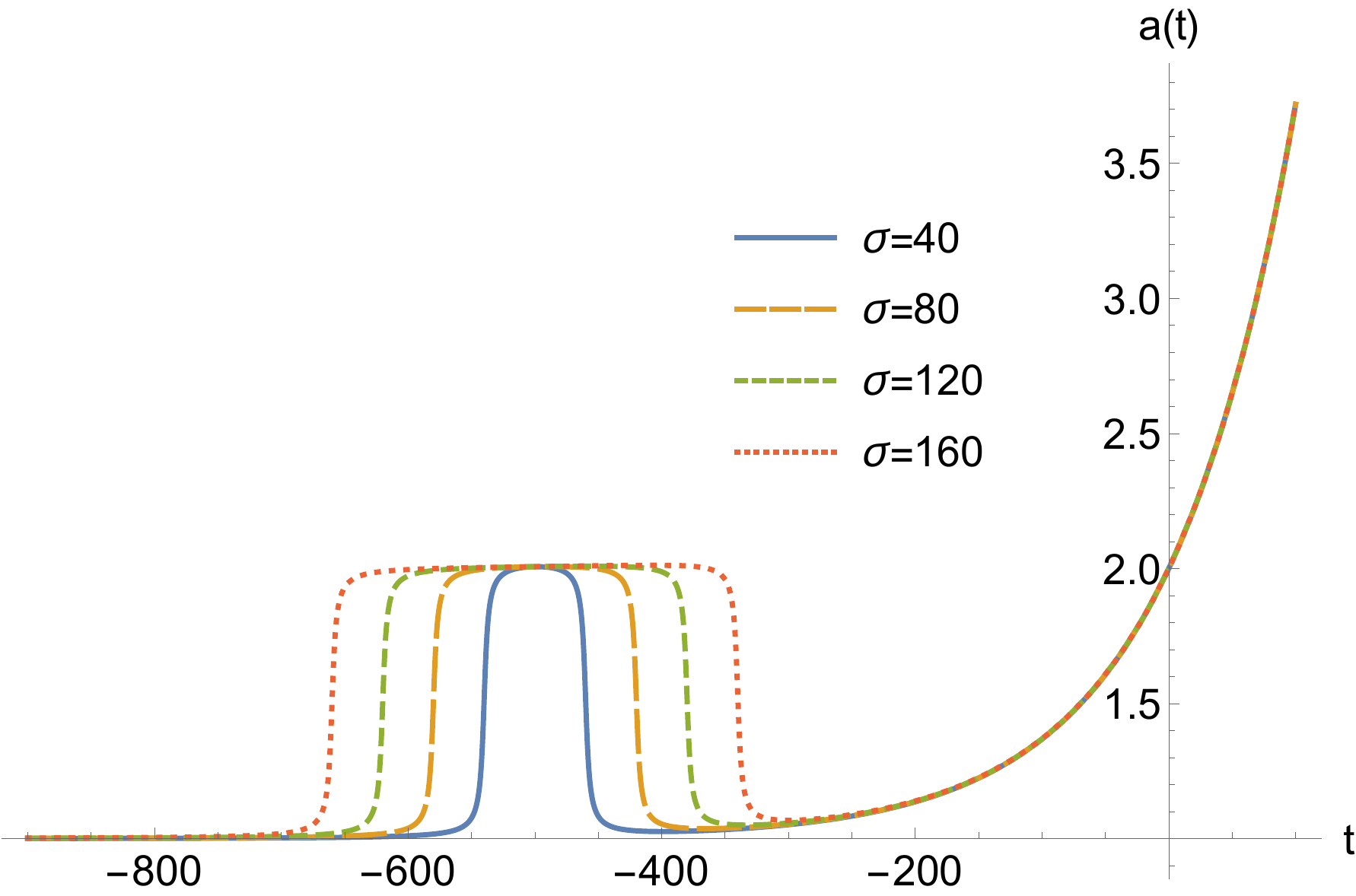}
  \end{minipage}
  \begin{minipage}[t]{0.4\textwidth}
     \includegraphics[scale=0.40]{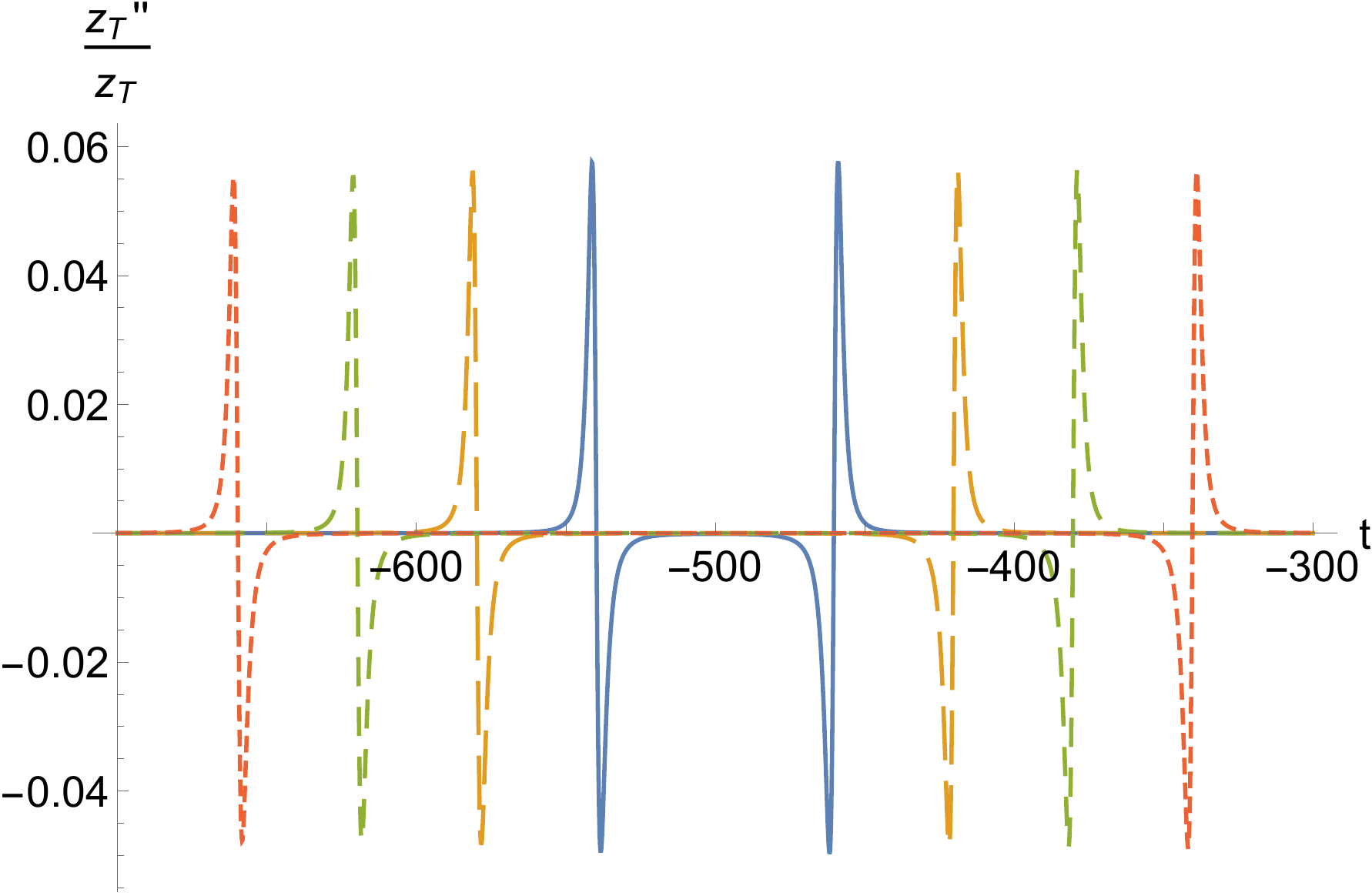}
  \end{minipage}
   \caption{\underline{Upper panel:} Evolution of the scale factor with a bounce centered on $\mu = -500$ and different values of $\sigma$. The other parameters of the model are unchanged with respect to the reference case. \underline{Lower panel:} Associated tensor potentials.}  
   \label{different Sigma}
\end{figure}

We turn to the study of the bounce width. The considered variations and their consequences on the tensor potential are shown in Fig. \ref{different Sigma}.\\

\begin{figure}[!tbp]
  \centering
  \begin{minipage}[t]{0.4\textwidth}
   \includegraphics[scale=0.37]{referenceSpectrum.pdf}
  \end{minipage}
  \begin{minipage}[t]{0.4\textwidth}
     \includegraphics[scale=0.37]{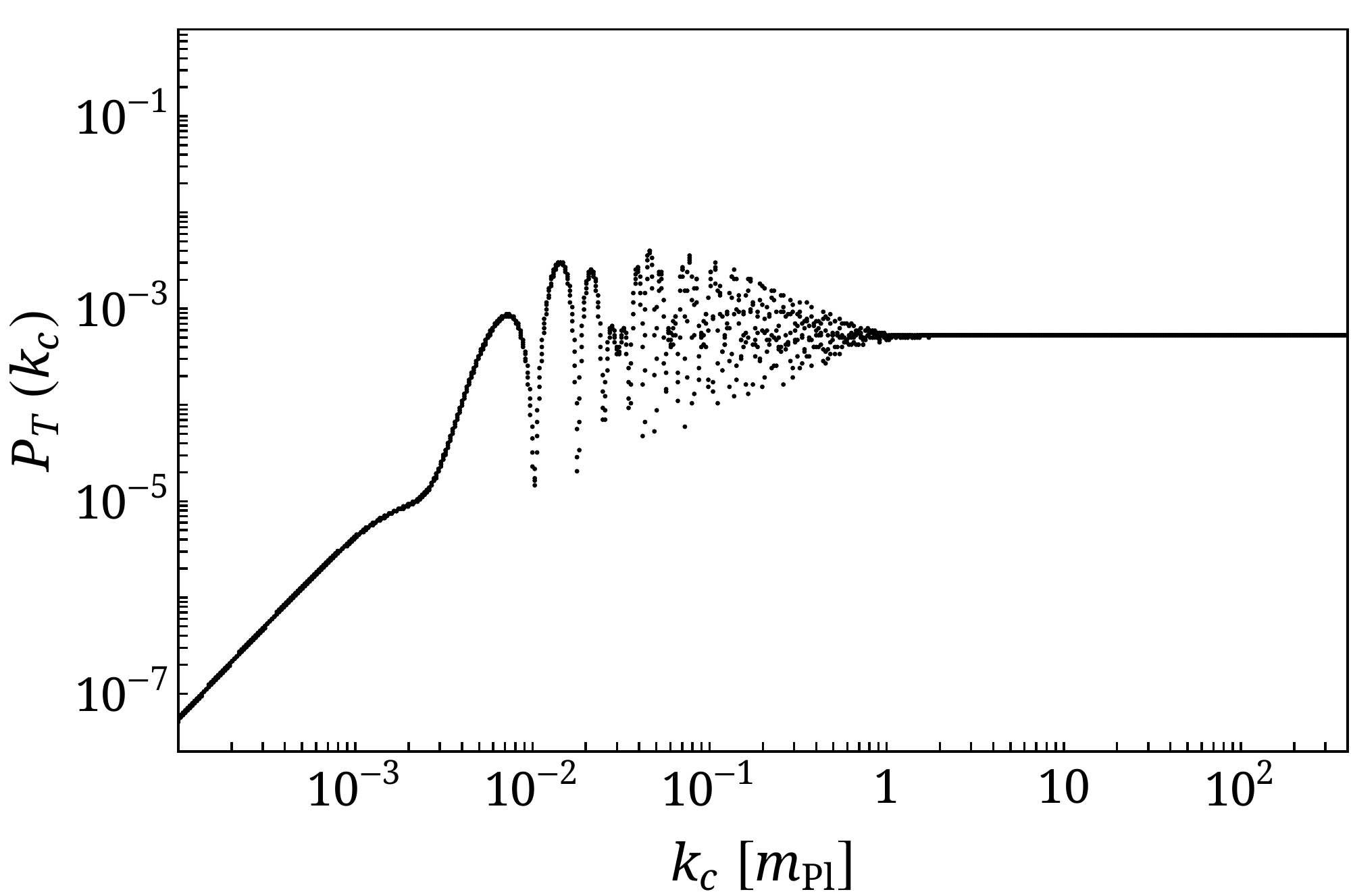}
  \end{minipage}
   \caption{\underline{Upper panel:} Spectrum of reference. \underline{Lower panel:} Spectrum with a wider bounce described by $\sigma=100$ and shifted to $\mu = -500$, the other bounce parameters remaining unchanged with respect to the reference case.}  
  \label{Different Sigma Spectrum}
\end{figure}

The impact of a modification of the width of bounce on the primordial tensor spectrum is shown in Fig. \ref{Different Sigma Spectrum}. For clarity and without any explicit consequence, the bounce position has been shifted to $\mu=-500$. The values of $\sigma$ are varied, the amplitude and the steepness remaining, as usual, unchanged. The main characteristics of the power spectrum are not significantly affected. Unless compensating for the steepness variation, as explained previously, the width of the bounce is unlikely to produce any spectacular observational consequence. 

\subsection{Impact of the parameters non related to the bounce}

\subsubsection{\textbf{The Hubble parameter during inflation $H_{0}$}}
 
\begin{figure}[!h]
\begin{center}
\includegraphics[scale=0.40]{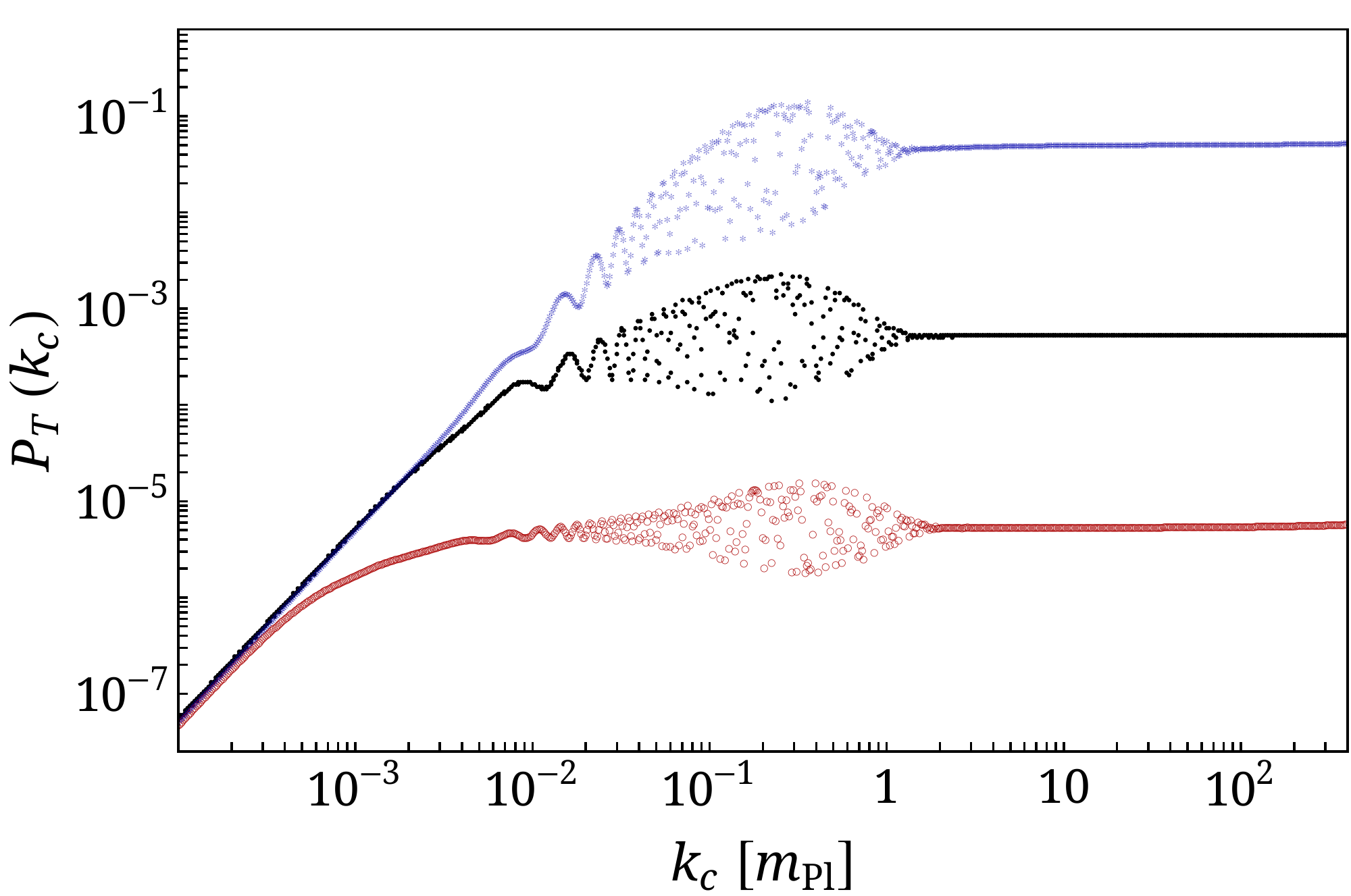}
\caption{Primordial tensor power spectra obtained from an evolution with a bounce characterized by $\mathcal{C} = 1$, $\mu =-400$, $\sigma = 2$ and $B=0.4$, and three different values of the Hubble parameter during inflation. From top to bottom: $H_{0}=0.1$ (blue stars), $H_{0}=0.01$ (reference case, black disks) and $H_{0}=0.001$ (red circles).} 
\label{Spectra different H0}
\end{center}
\end{figure}

 In this section we focus on the consequences of the inflationary stage on the tensor spectrum. As well known, a long enough inflationary phase leads, when combined with an appropriate choice of initial vacuum, to a scale invariant spectrum. We have varied $H_{0}$ and studied the impact on the spectrum. The results for $H_{0}=0.01$, $H_{0}=0.1$ and $H_{0}=1$ are given in Fig. \ref{Spectra different H0}.\\

Changing the value of $H_{0}$  modifies the amplitude of the power spectrum in the ultraviolet regime. More precisely, exactly as expected, the power is proportional to $H_{0}^{2}$, as in standard cosmology. 
Varying the value of the Hubble parameter does not change the picture beyond any standard and expected effect.

\subsubsection{\textbf{The normalization of the scale factor}}

\begin{figure}[!h]
\begin{center}
\includegraphics[scale=0.40]{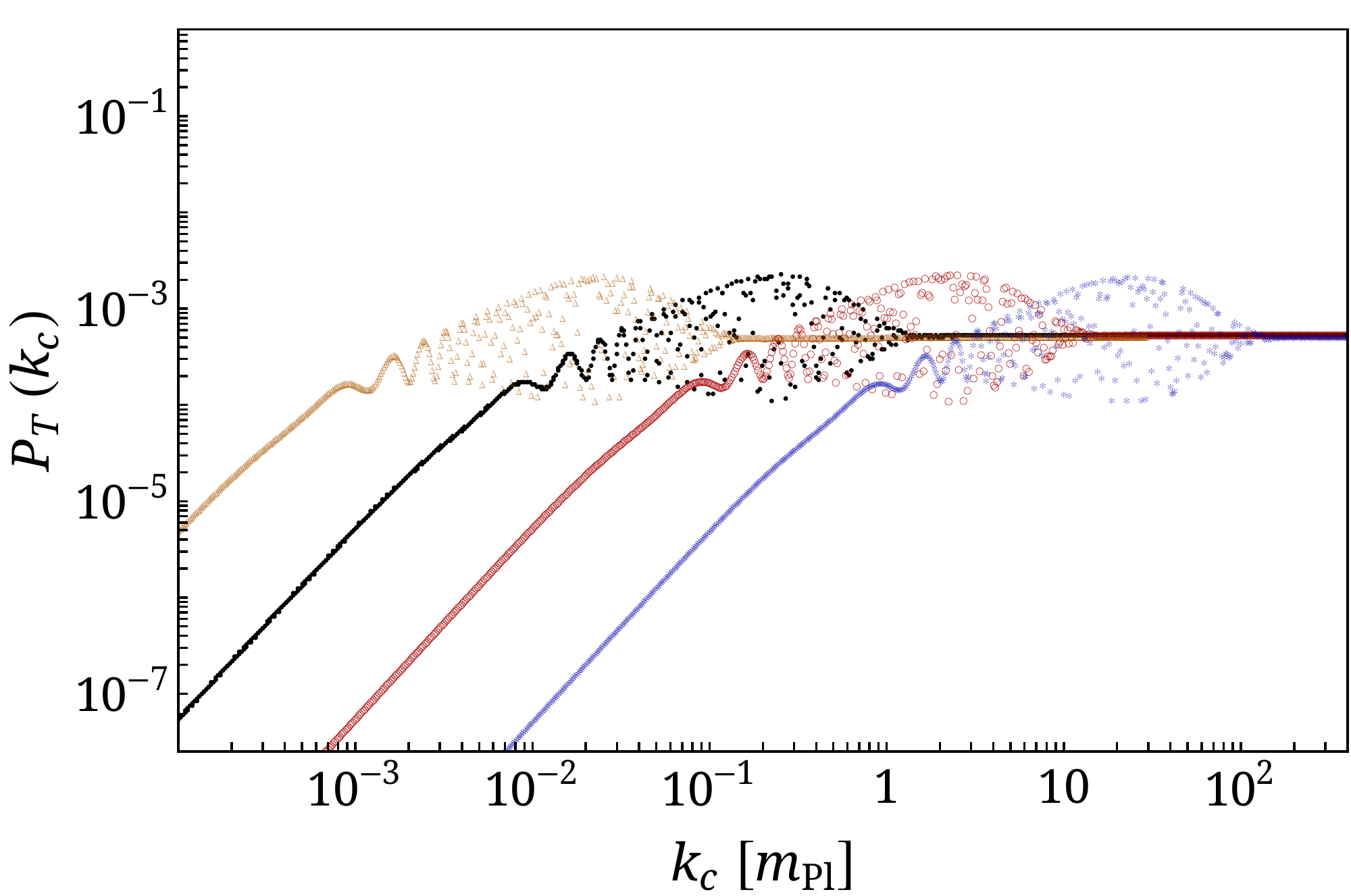}
\caption{Primordial tensor power spectra obtained from an evolution with a bounce characterized by  $\mathcal{C} = 1$, $\mu =-600$, $\sigma = 2$ and $B=0.4$, and different values of $A$. From left to right: $A=0.1$ (orange triangles), $A=1$ (reference case, black disks) and $A=10$ (red circles) and $A=100$ (blue stars).} 
\label{Spectra different A}
\end{center}
\end{figure}

In this section,  $H_{0}$ is kept fixed to $H_{0}=0.01$ and we investigate the impact of variations of the constant $A$. In principle, this is just an unphysical  rescaling of the scale factor. However, the case of effective quantum cosmology is slightly more subtle as an extra fundamental scale (presumably of the order of the Planck length) might enter the game. This is not the case in the toy model we consider here but this is clearly the case in quantum reduced loop gravity \cite{Alesci:2016xqa,Alesci:2018qtm}. In this situation, the conjugate variables ($a$ and $H$) should not be understood as describing the Universe as a whole but instead as referring to a fundamental (or elementary) cell \cite{Barrau:2014maa,Bojowald:2015iga}. To make simple the use of our results in another context we therefore present in Fig. \ref{Spectra different A} the effects of a variation of the constant $A$. As expected, a rescaling of the scale factor just shifts the spectrum such that the physical wavenumbers values remain unchanged.
 
\section{Multiple bounces}

It is worth considering, in addition to the first reference bounce, another perturbation of the scale factor in the static phase, that is a scale factor given by:

\begin{eqnarray}
& & a(t)  =  A + A e^{H_{0} t} + A \frac{C}{2 \arctan\left(B \sigma \right)} \times \\ \nonumber
&& \left\lbrace \arctan \left[ B \left( t-\left( \mu - \sigma \right) \right) \right] - \arctan\left[B \left(t-\left(\mu + \sigma \right) \right) \right] \right\rbrace \\ \nonumber
&& + A \frac{C^{\star}}{2 \arctan\left(B^{\star} \sigma^{\star}\right)} \left\lbrace \arctan \left[B^{\star} \left( t-\left( \mu^{\star} - \sigma^{\star} \right) \right) \right]  \right. \\ \nonumber 
&& \left. - \arctan\left[B^{\star} \left(t-\left(\mu^{\star} + \sigma^{\star} \right) \right) \right] \right\rbrace ~,
\end{eqnarray}

in which the parameters labelled with the ``$^{\star}$" symbol are the analogous of $C$, $B$, $\mu$ and $\sigma$ for the new additional bounce. Once again, although this possibility is in principle generic and fully phenomenological it is also motivated by some quantum gravity results.\\
\begin{figure}[!h]
\begin{center}
\includegraphics[scale=0.42]{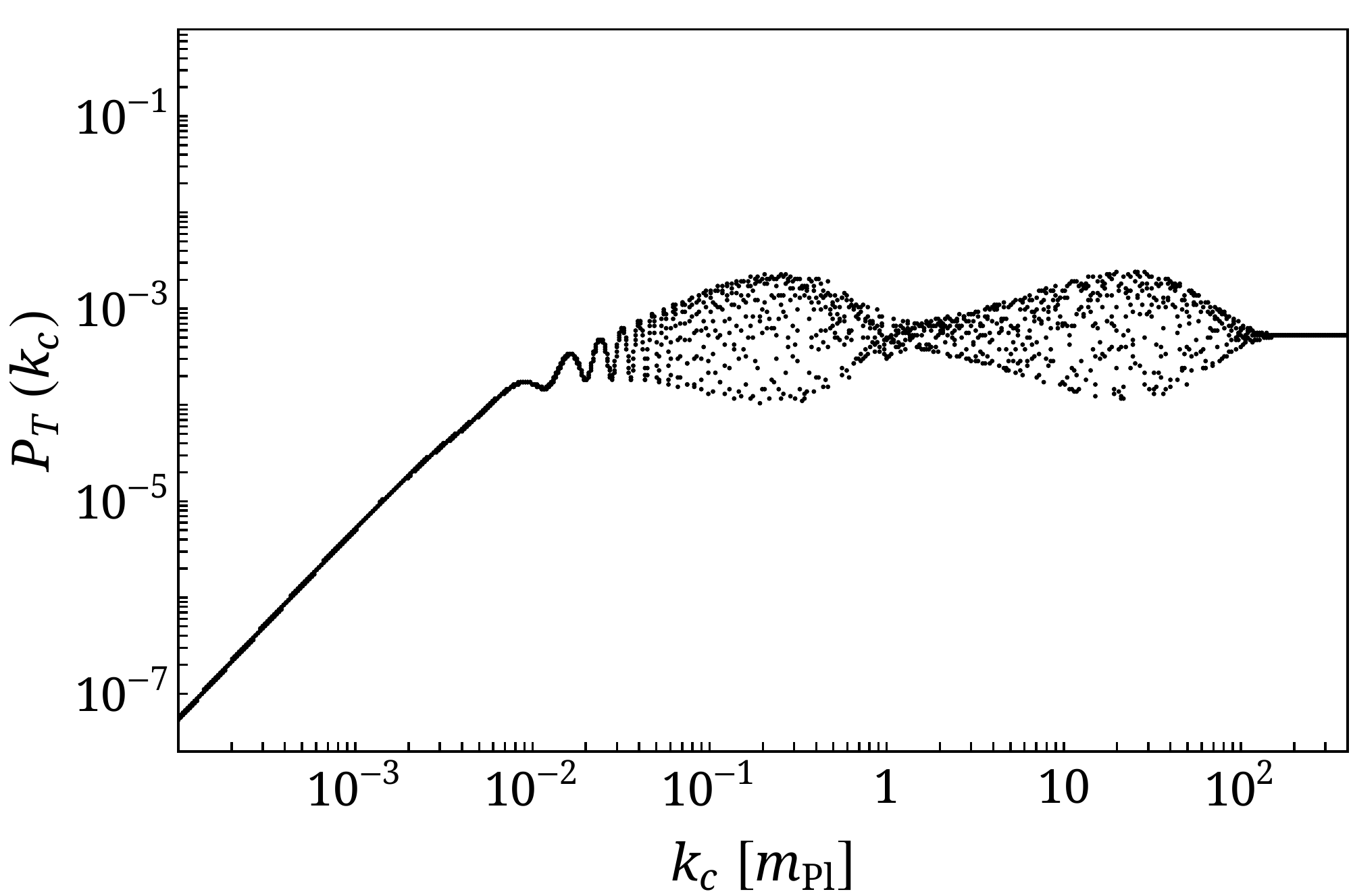}
\end{center}
\caption{Primordial tensor power spectrum obtained from an evolution with two bounces of different steepness, the first one being described by $\mathcal{C}=1$, $\mu =-400$, $\sigma = 2$ and $B=0.4$, and the second by $\mathcal{C}^{\star}=1$, $\mu^{\star} =-400$, $\sigma^{\star} = 2$ and $B^{\star}=40$. The Hubble parameter during inflation is $H_{0}=0.01$. }
\label{spectra multiple bounces}
\end{figure}

The spectrum corresponding to an evolution with two bounces of different steepnesses is shown in Fig. \ref{spectra multiple bounces}. One can notice the presence of two ``bullet" features, one for each bounce of the scale factor. It is possible to vary the characteristics of each bounce, thus the characteristics of each "bullet", independently by adjusting appropriately the parameters. If the two bounces have the same width, even if their positions are far one from the other, then the two bumps are perfectly superposed in the spectrum. If, however, the shapes of the bounces differ, they might be distinguishable in the tensor spectrum and observational footprints of the details of the transition phase might be expected.

\section{Conclusion}

In this article we have clarified some general properties of the primordial cosmological tensor power spectrum in emergent models. Following a purely phenomenological approach, we have studied how different features in the behavior of the scale factor around the transition time (or before) can affect the spectrum. The main result are the following:
\begin{itemize}
\item in itself, the existence of a static phase in the remote past of the Univers does not lead to a scale invariant power spectrum. 
\item if the static phase is followed by a long enough stage of inflation, the spectrum might become flat in the observable range of wavenumbers.
\item the consequences of the details of the evolution of the scale factor around the transition time, modeled as a mini-bounce (or antibounce), are not erased by inflation and appear as a ``bullet" feature in the spectrum.
\item the position of the mini-bounce has only a small influence on the shape of the "bullet" but its the steepness and his amplitude control respectively the comobile position and the size of the bullet.
\item multiple bounces can leave complex features in the spectrum. Bounces with different characteristics might leave distinguishable imprints in the tensor spectrum.
\end{itemize}

This work  establishes that non-trivial features occurring at the transition time in an emergent universe might be detectable in the promordial tensor spectrum. The detection of the CMB B-modes is a very active field involving big collaborations. On the ground, progresses are expected from  BICEP or POLARBEAR (now grouped into Stage 4) exepriments, and, in space, potentially from LiteBIRD. At this stage, trying to detect those modes is probably the best path toward finding traces of quantum gravity effects in the CMB. The features studied in this work may therefore be observable in a not so far away future, if the duration and energy scale of inflation are favorable.

It would clearly be interesting to go beyond the tensor spectrum and to investigate scalar perturbations that are currently observed. This, however, requires an explicit specific model as the evolution of the scale factor is not anymore enough to compute the evolution of perturbations.

\section{Acknowledgments}

K.M is supported by a grant from the CFM foundation.

\bibliography{refs}

\end{document}